\documentclass[10pt,journal]{IEEEtran}

\ifCLASSOPTIONcompsoc
  \usepackage[nocompress]{cite}
\else
  \usepackage{cite}
\fi
\ifCLASSINFOpdf
\else
\fi
\usepackage{comment}
\usepackage{graphicx}
\usepackage{array}
\usepackage{graphics}
\usepackage{pdfpages}
\usepackage{amsmath,amssymb,amsfonts}
\usepackage{algorithmic}
\usepackage{graphicx}
\usepackage{array}
\usepackage{graphics}
\usepackage{textcomp}
\usepackage{comment}
\usepackage{rotating}
\usepackage{adjustbox}
\usepackage{lscape}
\usepackage{hyperref}
\usepackage{amsmath}
\usepackage{multirow}
\usepackage{graphicx}
\usepackage{xcolor}
\usepackage{color,soul}
\usepackage{multirow}
\usepackage{graphicx}
\begin{document}
\title{Multimodal Driver State Modeling through Unsupervised Learning}

\author{Arash~Tavakoli, and Arsalan Heydarian 
\IEEEcompsocitemizethanks{\IEEEcompsocthanksitem Arash Tavakoli and Arsalan Heydarian (corresponding author) are with the Department
of Engineering Systems and Environment, University of Virginia,
VA, 22901.\protect\\
E-mail: ah6rx@virginia.edu
}

}

\IEEEtitleabstractindextext{%
\begin{abstract}

Naturalistic driving data (NDD) can help understand drivers' reactions to each driving scenario and provide personalized context to driving behavior. However, NDD requires a high amount of manual labor to label certain driver's state and behavioral patterns. Unsupervised analysis of NDD can be used to automatically detect different patterns from the driver and vehicle data. In this paper, we propose a methodology to understand changes in driver's physiological responses within different driving patterns. Our methodology first decomposes a driving scenario by using a Bayesian Change Point detection model. We then apply the Latent Dirichlet Allocation method on both driver state and behavior data to detect patterns. We present two case studies in which vehicles were equipped to collect exterior, interior, and driver behavioral data. Four patterns of driving behaviors (i.e., harsh brake, normal brake, curved driving, and highway driving), as well as two patterns of driver's heart rate (HR) (i.e., normal vs. abnormal high HR), and gaze entropy (i.e., low versus high), were detected in these two case studies. The findings of these case studies indicated that among our participants, the drivers' HR had a higher fraction of abnormal patterns during harsh brakes, accelerating and curved driving. Additionally, free-flow driving with close to zero accelerations on the highway was accompanied by more fraction of normal HR as well as a lower gaze entropy pattern. With the proposed methodology we can better understand variations in driver's psychophysiological states within different driving scenarios. The findings of this work, has the potential to guide future autonomous vehicles to take actions that are fit to each specific driver.

\end{abstract}

}

\maketitle

\IEEEdisplaynontitleabstractindextext
\IEEEpeerreviewmaketitle

\ifCLASSOPTIONcompsoc
\IEEEraisesectionheading{\section{Introduction}\label{sec:introduction}}
\else


\section{Introduction}
\fi
People have different preferences in their choice for driving styles \cite{park2020driver}. For instance, a recent study mentioned individual variability across participants when choosing between conservative and risky driving styles \cite{park2020driver}. Additionally, different studies have shown that personalization of autonomy has a positive effect on the user's perceived trust in autonomous vehicles (AV) \cite{motamedi2020acceptance,sun2020exploring,zhang2021drives,xing2021toward}. This implies that AV's acceptance relies on correctly understanding people's preferences in driving styles and act accordingly. One method to identify individual responses to different driving behaviors is using historical driving data. For instance, driver-related physiological data (e.g., facial expressions, heart rate (HR), and skin temperature) can show how a driver reacts within different driving styles \cite{yamakoshi2009novel,katsis2011wearable,nacpil2021application}.

Analysis of historical driving data, which is often retrieved through naturalistic driving studies (NDS), also comes with problems such as the difficulty (time and cost) in analyzing the massive amount of collected information \cite{li2021extraction}. NDS is often conducted in a longitudinal fashion to help detect behaviors while different environmental noise and challenges exist in the data \cite{tavakoli2021harmony}, which drastically increases the amount of data that needs to be analyzed for detecting specific driver behaviors, actions, and responses. One method to address this issue is to apply unsupervised learning on both driving behaviors as well as the driver's state. Through unsupervised learning, different patterns in driver's state and behaviors can be detected \cite{li2021extraction}. These patterns, when analyzed together, may reveal important information regarding the driver's  responses in each driving behavior.

Previous NDS data often lack information related to driver's physiological responses and cognitive metrics \cite{carsten2013vehicle,tavakoli2021harmony}. With the current advancements in wearable technology, it is now viable to detect driver's psychophysiological states through monitoring their heart rate (HR), skin temperature, skin conductance, arm movement, and other physiological metrics. In line with these improvements, recent studies have used driver's physiological responses such as changes in HR to detect driver's stress levels correlated with the increase in HR values \cite{tavakoli2021harmony, zepf2019towards}. Additionally, with the improvements in computer vision applications, it is now viable to use raw videos for estimating driver's breathing patterns, HR, gaze directions, and emotional features through facial landmarks \cite{perepelkina2020hearttrack}. As a result of these improvements in technology, recent NDS are starting to implement novel sensors and technologies to collect such data \cite{farah2012evaluation,zepf2019towards,qiu2019analysis,tavakoli2021harmony,fridman2019advanced}. Utilizing these data streams and by applying unsupervised machine learning techniques, we can understand a driver's response within each driving behavior. 

In this paper, we propose a methodology to classify driver's reactions to different vehicular maneuvers using unsupervised learning in a fully naturalistic environment. We use a case study to analyze different driving patterns (e.g., harsh braking) coupled with driver state patterns (e.g., high stress level) extracted from a naturalistic driving dataset in an unsupervised fashion. Using the detected unsupervised patterns and by having access to multimodal human sensing data, we then analyze the driver's reactions under each driving pattern. Specifically, we first decompose naturalistic driving scenario data into smaller segments using Bayesian Change Point (BCP) detection methods. We specifically use BCP for its ability to detect changes in the distribution of multimodal data resulting from vehicle kinematic sensors. Then, a combination of Gaussian Mixture Models (GMM) together with Latent Dirichlet Allocation analysis (LDA) is used to provide patterns of both driving behavior and driver state. We use the patterns in driver state to understand driver reactions with respect to specific driving behaviors. In order to show the applicability of our approach, we use two separate case studies. In the first one, we use high-frequency driving behavior (e.g., acceleration) and driver sensing data (i.e., HR and gaze) from one participant and apply the GMM-LDA method on both driving behavior and driver sensing data. Based on this model, we detect four driving patterns that are aligned with well-known driving behaviors of harsh braking, normal braking, highway free-flow driving, and road curvature driving. We also detect two driver state patterns of abnormal (high) HR versus baseline HR, as well as low versus high gaze entropy. Analyzing the fraction of each state pattern (i.e., HR and gaze pattern) within each driving behavior depicts that the driver of our study is calmer in a more conservative driving style with route selection that can include more segments of highway driving. We then extended our findings to a larger dataset with 12 participants as a second case study with low-resolution data and found similar results as case study I.

\section{Background Study}\label{sec:background}
Previous studies have pointed out the utility of unsupervised learning methods in finding behavioral patterns in naturalistic driving data \cite{li2021extraction}. However, past studies have not focused on applying the same methods on driver physiological data (e.g., HR). Below we first elaborate on the relationship between driver sensing measures (i.e., HR and gaze patterns) and driver's state (i.e., stress level, and workload). We then outline past progress in unsupervised modeling techniques for driver state and behavior detection. 


\subsection{Driver State Detection}\label{sec:state}
Human bio-signals have been used extensively in multiple fields such as psychology, health sciences, and engineering for retrieving a deep understanding of a human's psychophysiological states. Psychophysiology refers to psychological states such as emotional responses (e.g., anger, frustration, and happiness), cognitive load, and distraction, which can be measured through changes in human physiological responses (e.g., HR, skin temperature, and skin conductance) \cite{lohani2019review}. In the driving research, psychophysiological measures such as driver's HR \cite{sugie2016detection,tavakoli2019multimodal,tavakoli2021leveraging}, gaze patterns \cite{baee2019medirl,tavakoli2021harmony}, skin conductance \cite{pakdamanian2020deeptake}, and brain signals \cite{pakdamanian2020toward} were all used for retrieving driver's stress level, and cognitive load.  

One popular definition of stress is the state of a human when the demand of a situation is perceived to be more than the available internal resources \cite{francis2018embodied}. Different bio-signals can then be used to detect stressful instances. Human cardiovascular measures have been extensively used in literature for detecting stress levels. In naturalistic environments and mainly driving research, cardiovascular activity can be retrieved through using either of Electrocardiography (ECG) or photoplethysmogram (PPG) technologies. ECG measures heart electrical activity through the usage of contact electrodes. PPG, on the other hand, records the same activity through measuring blood volume in the vein using infrared technology \cite{lohani2019review,tavakoli2021harmony}. The cardiovascular measures can then be used to estimate features of the beat-to-beat signal of the heart such as HR itself, and root mean squared of successive intervals (RMSSD). Studies have demonstrated that higher stress levels are correlated with an increase in HR and a decrease in RMSSD \cite{tavakoli2021harmony,kim2018stress}.


Driver's cognitive workload is defined as cognitive resources that are taken from the driver by any activity other than the driving itself \cite{engstrom2017effects}. In this definition, the cognitive load consist of mind wandering and the load imposed on the driver by secondary tasks. Secondary tasks are the ones that require attention but are not required for the task of driving, such as working with a phone \cite{engstrom2017effects}. Human bio-signals such as driver's gaze measures, cardiovascular measures, and brain signals have been used for cognitive load estimation in both controlled environments and in naturalistic settings \cite{lohani2019review}. Stationary and transition gaze entropy are two of the main eye gaze metrics that are commonly used for cognitive load estimation \cite{krejtz2018eye,fabio2015influence,shiferaw2018stationary,shiferaw2019review}. In information theory, the uncertainty associated with a choice is referred to as the information entropy \cite{shiferaw2019review} in which the more the uncertainty, the more the entropy and the more randomness in the system. There are two ways to calculate the entropy in gaze analysis. The first is generally calculated through Shannon's equation \cite{shannon1948mathematical}. In gaze analysis, Shannon's entropy shows the overall predictability of fixation locations in a sequence of gaze patterns, which is a measure of gaze dispersion \cite{shannon1948mathematical} and is referred to as the Stationary Gaze Entropy (SGE). Specifically, for a set of fixation locations in a sequence of eye movements, fixation locations can be assigned to spatial bins of $p_i$ and calculate the SGE as:
\begin{equation}
SGE = -\sum_{\textit{i=1}}^{n} p_{i} \log_{2}p_{i} 
\end{equation}

Different studies used SGE to infer a human's state in different conditions. For instance, SGE has been shown to be correlated with task difficulty, complexity, and experience with the task \cite{shiferaw2019review}. Additionally, studies have used SGE in driving research to infer driver's state such as workload \cite{shiferaw2019review}, drowsiness \cite{shiferaw2018stationary}, and being under the influence of alcohol \cite{shiferaw2019gaze}. For instance, \cite{shiferaw2018stationary} used driver's eye SGE to predict lane change events in sleep-deprived drivers, where an increase in SGE was associated with a higher probability of lane change events. While SGE was shown to be correlated with different human states, a recent review suggests that inferences based on changes in SGE can be very task-specific. For instance, if we know a specific task requiers a higher SGE and we observe that the participant is having a lower SGE, this may imply the participant could be disengaged from the task \cite{shiferaw2019review}. This highlights the importance of the second measure of gaze entropy, referred to as conditional entropy, which is task-independent. 

Conditional entropy considers the dependency between back-to-back fixation points in a temporal fashion. This results in the Gaze Transition Entropy (GTE). GTE is a measure of predictability of the next fixation location given the current location. For a sequence of transitions between different spatial bins of $i$ and $j$, with a probability of $p_{ij}$, the GTE is calculated as:
\begin{equation}
GTE = -\sum_{\textit{i=1}}^{n} p_{i} \sum_{\textit{j=1}}^{n} p_{ij} \log_{2}p_{ij} 
\end{equation}

GTE was shown to be correlated with multiple aspects of the human's state in both driving and non-driving research. In general, higher task demand, higher scene complexity, and higher levels of cognitive load were shown to increase the GTE \cite{shiferaw2019review,fabio2015influence}. Additionally, higher levels of GTE when performing the same task with the same scene complexity can be associated with higher stress levels \cite{shiferaw2019review}.

\subsection{Unsupervised Modeling of Driver's state and Behavior}
Previous studies provided significant insight into analyzing driver's behaviors in naturalistic conditions through unsupervised methods. Bando et al. \cite{bando2013automatic} have used multimodal LDA to infer driving topics using a combination of image sequences, annotated tags and driving behavioral data. In their study, authors first used a double articulation analyzer (DAA) for driving data segmentation. Then by applying LDA on the multimodal data, they were able to achieve a dimensionality reduction of 5\% on the raw data. This then led to an increase in classification accuracy achieved by a baseline that was trained using a support vector machine. In another study, the authors proposed a modification to DAA to not only perform segmentation but also to predict the duration of each segment \cite{taniguchi2014unsupervised}. Based on the assumption that driving data has a two-layer hierarchical structure, the authors proposed a double-layer articulation structure model which uses a hierarchical Dirichlet process hidden semi-Markov model to predict the duration of each segment. Li et al. \cite{li2016driver} used the density-based spatial clustering method to cluster physiological data of drivers into one of the normal, event, and noise clusters. Based on the data collected from three drivers and through an on-road controlled study, authors were able to achieve a recall rate of 75 percent in clustering the three categories.

Bender et al. \cite{bender2015unsupervised} proposed an unsupervised method to provide high-level clusters for time series data streams from naturalistic driving behavior. In their study, the authors used a Bayesian multivariate linear model to segment the driving data. Then by using simplicial mixture models, they find unsupervised patterns associated with different driving behaviors such as acceleration and braking. This method was applied in an online fashion through an on-road controlled study. The data for the study was collected from a vehicle equipped with multiple telemetry sensors driving a 13 min route. Wang et al. \cite{wang2018learning} proposed a method to build a library of human's basic driving motion primitives that can be used for human-like actions. In their paper, the authors have used a probabilistic inference based on iterative expectation-maximization (EM) algorithm on driving data collected from one driver. Their probabilistic method achieves a more meaningful segmentation when compared with the classical EM-GMM approach by merging back-to-back clusters together, which is closer to the real-world driving situation. Lastly, \cite{li2021extraction} proposed a framework to automatically provide a description for driver's behavioral data retrieved through telemetry sensors inside the vehicle. In their method, they applied Bayesian multivariate linear regression to segment the driving data. Through using three different clustering methods of Gaussian mixture model LDA (GMM-LDA), Gaussian Wishart LDA (GW-LDA), and Multimodal LDA (mLDA), they found out that GMM-LDA provides the most useful description generation for naturalistic driving behavior data.

Although previous studies have provided significant insights into the application of unsupervised learning in driver states and behavior detection, most of them did not couple the two together to understand the driver's state in each behavioral pattern. This is extremely important as combining driver's states with behavioral patterns can provide a deeper description for each driving segment. For instance, while the unsupervised category for a driving segment might be ``high acceleration'', it is important to know whether the driver was carrying a high work load or not in that segment. For the future AV, such information might help with better prediction of driver's state in each driving scenario. Additionally, most of the past studies in driver state detection are performed in experimental conditions either in a driving simulator or in an on-road controlled study where the conditions are different from a fully natural environment. Such shortcomings mostly existed due to not having access to naturalistic human sensing data. This is now partly achieved through novel wearable technology devices that are implemented day after day in driving studies (see \cite{tavakoli2021harmony,zepf2019towards}. In this paper, by performing an unsupervised analysis on both driver state data collected through conventional wearable devices and driver's behavioral data collected through vehicle's kinematic data, we find driver's state (i.e., stress levels and workload) in each driving pattern.

\section{Methodology}

In this section the framework for retrieving state and behavior patterns is described. We then apply our framework to a fully naturalistic driving data collected through one of our previous studies \cite{tavakoli2021harmony}. We first perform a data exploration using our framework on a driver's behaviors and states extracted through a 2 hour and 10 minutes long trip of a vehicle equipped with multiple sensors collecting contextual information including both driver and environmental sensing modules (i.e., case study I). We then apply our method to a larger pool of data collected from 12 participants (i.e., case study II). Below we first outline the methodology followed by the dataset details, the selected parameters.

In our framework, kinematic Sensors are the ones that record movements and acceleration in different directions such as an inertial measurement unit (IMU), which records acceleration and angular velocity in 3 different directions of X, Y, and Z. These information are used as driving behavior data such as vehicle's lateral and forward acceleration. On the other hand, human sensing modules are sensors specifically used for human related data such as smart watches and in-cabin and outdoor facing cameras. These data are used to detect drivers state such as stress level and workload.   

The goal of this framework is to apply unsupervised learning methods on the data from \textbf{kinematic sensor} readings and \textbf{human sensing modules}, to detect \textbf{driving-related} and \textbf{driver state-related} patterns automatically. These patterns are then compared with  known \textbf{driving behaviors} (e.g., harsh acceleration) and \textbf{driver states} (e.g., high stress level) to provide descriptions for each detected pattern. We then statistically compare different driving behaviors based on the proportion of each driver state within them. 

\begin{figure*}
\begin{center}
  \includegraphics[width=0.95\linewidth]{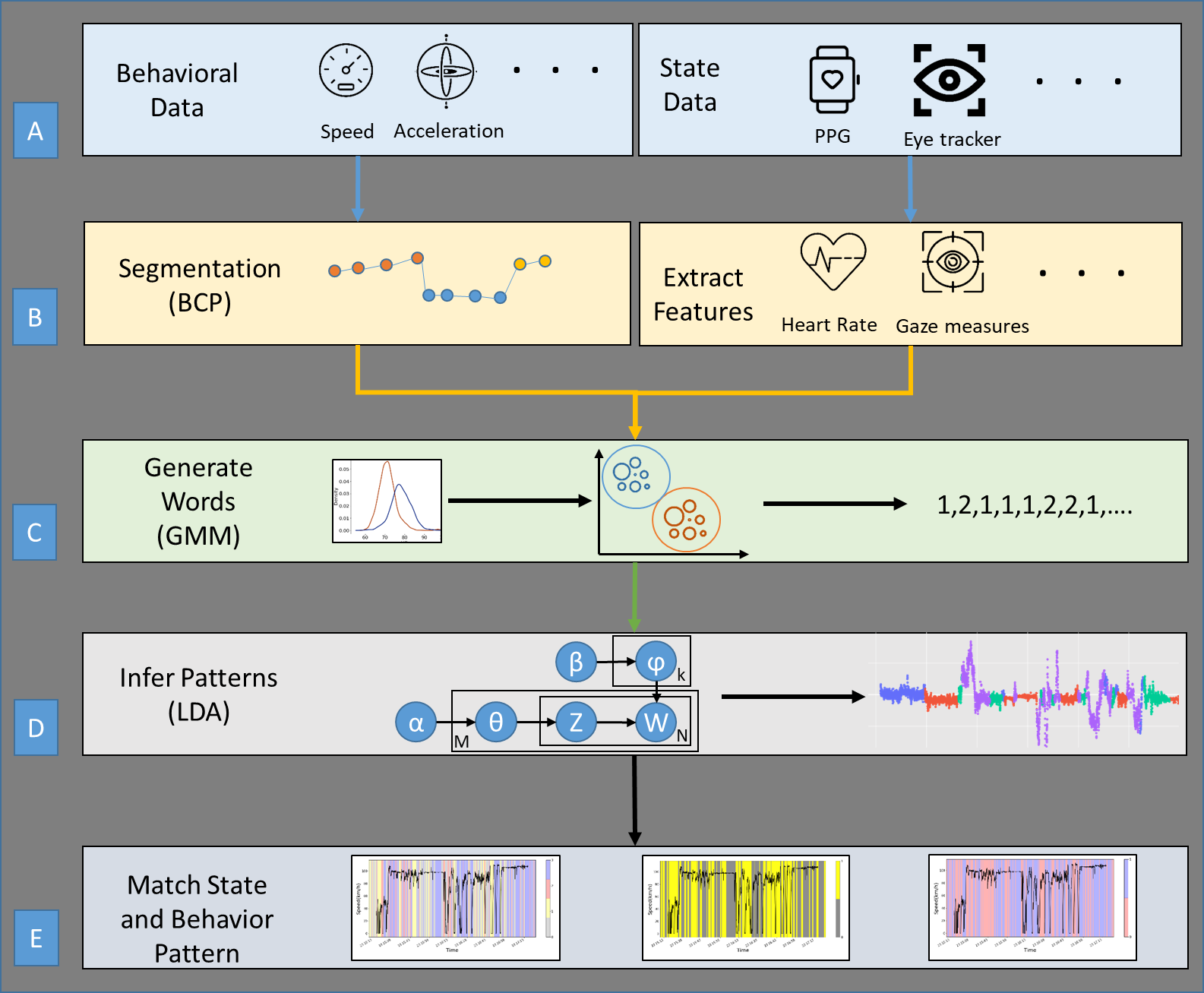}
  \end{center}
  \caption{An overview of the framework of analysis consisting of six main sections. At first we retrieve features from driver's state data that align with stress level and work load (A and B). Additionally, we perform segmentation on behavioral data that provides specific segments in driving (B). We then perform unsupervised analysis through GMM-LDA method to find different patterns in driver's behavior and state. We first apply GMM to find word-like objects in the continuous inputs (C). We then apply the LDA on the sequence of word like objects in each segment to find different patterns in state and behavior (D). Lastly, we assess the presence of each state pattern within in driving behavioral pattern to find the more suitable driving behavior pattern to the driver.}
  \label{fig:framework-analysis}
\end{figure*}

Our framework consist of six main sections. As a summary, through a formerly proposed NDS framework, multiple kinematic  and human sensing data-streams are collected (Fig. \ref{fig:framework-analysis} - A). Driver's stress level and work load information are retrieved from human sensing data (Fig. \ref{fig:framework-analysis} - B). Similarly, driving behavior data is used to retrieve driving segments (Fig. \ref{fig:framework-analysis} - B). Then through an unsupervised learning method, a driving behavior and a driver state pattern is generated for each driving segment (Fig. \ref{fig:framework-analysis} - C and D). Lastly, within each detected behavioral pattern we assess the likelihood occurrence of each state patterns (Fig. \ref{fig:framework-analysis} - E). 


\subsection{Details of Case Study Datasets}\label{sec:case_study_data}
\begin{figure*}
\begin{center}
  \includegraphics[width=0.95\linewidth]{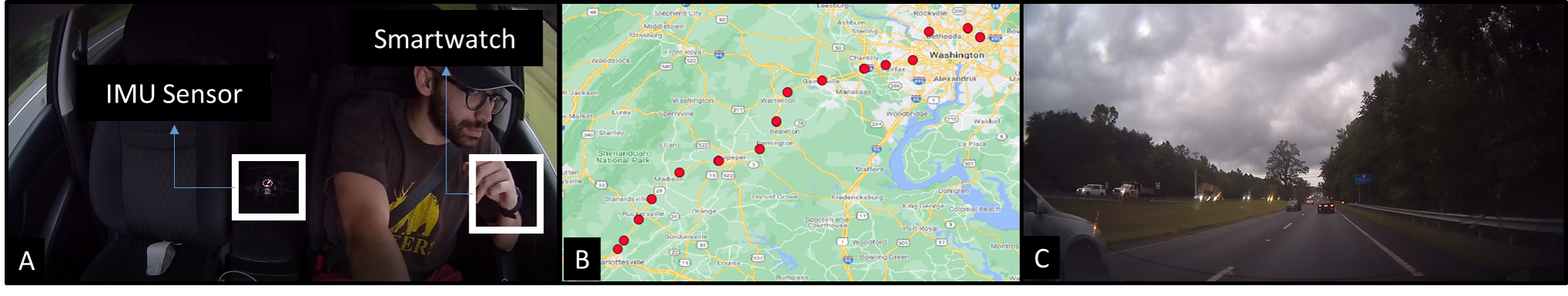}
  \end{center}
  \caption{An overview of the case study dataset. The vehicle is equipped with a camera, smart watch, and an IMU sensor (A, and C). The data is collected from a 120 mile driving scenario between cities of Charlottesville and Washington DC, with a duration of 2 hour and 10 minutes (B).}
  \label{fig:framework-data}
\end{figure*}

In our study we used two different datasets with different sensors and different resolutions to test the method. For case study I, we use the data collected from a 120 mile driving scenario between cities of Charlottesville and Washington DC. This data is available through an online repository \cite{Harmonydata}. Fig. \ref{fig:framework-data} shows the map of the trip. Based on the framework suggested in our previous work \cite{tavakoli2021harmony}, the vehicle has a dash camera which collects both in-cabin and outside environment as well as vehicle's speed. 

The videos collected from the in-cabin camera is fed into the OpenFace software \cite{baltrusaitis2018openface} to retrieve driver gaze patterns at 30 frame per second resolution. OpenFace can perform facial landmark detection and gaze direction analysis on one or multiple people within a frame. OpenFace uses Conditional Neural Fields (CLNF) \cite{baltrusaitis2013constrained} for facial landmark detection. As an open source off-the-shelf software, it has achieved the least error in landmark detection, gaze estimation and head pose estimation benchmarks. Specifically, it has achieved an error of 9.96 \% on the MPIIGAZE dataset \cite{zhang2015appearance}.   

Additionally, the driver is provided with a smartwatch that collects HR data, hand movement (through IMU sensor), and GPS locations. Also, we used an IMU sensor in the vehicle that collects vehicle's kinematic data including acceleration, linear acceleration as well as rotational velocity in 3 different directions. The detail of data collection and setup is provided in \cite{tavakoli2021harmony}.

In case study II, we use driving data collected from 12 participants through our previous study, HARMONY \cite{tavakoli2021harmony}. From each participant, we randomly chose a continuous highway trip that lasted approximately 90 minutes. In HARMONY, we did not have access to IMU sensors inside the vehicle. Instead, we calculated vehicle's forward acceleration through applying a gradient on the speed data retrieved from GPS on the camera. Although this GPS information was sampled at a much lower frequency (i.e., 1 HZ), it provides significant insights into differences across participants. 

\subsection{Driver State Feature Extraction}
We use driver's raw HR directly for driver stress inference as previous studies showed the positive correlation between driver's HR and stress levels (see section \ref{sec:state}, and Fig. \ref{fig:framework-analysis} - B). Moreover, our previous study in driving environments showed that stressful events on the road might change the distribution of driver's HR momentarily, and increase the HR for a short period of time \cite{tavakoli2021harmony}.  


As described in section \ref{sec:state}, driver's gaze patterns can be used for workload estimation (Fig. \ref{fig:framework-analysis} - B). For this purpose, from the driver's gaze data, we retrieve the GTE. For this task, we first construct a 2D space of the range of the gaze angles \cite{krejtz2015gaze,tran2020you}.  The 2D space is divided into equally distanced areas of interest (AOI) which can be a $4*4$ grid. The GTE is then calculated based on a rolling window of size 240s on the driving data \cite{krejtz2015gaze}. As a summary, for a sequence of AOIs, we find the transition matrix between AOIs by assuming a first-order Markov process for the gaze sequences \cite{krejtz2015gaze}. To this end, a transition matrix is retrieved with $p_{ij}$ being the probability of switching between AOIs i and j (in S), with a stationary probability of $\pi_i$. Then the GTE is calculated based on Shannon's entropy as:

\begin{equation}
\hat{H}_{t} = -\sum_{\textit{i} \in S} \pi_{i} \sum_{\textit{j} \in S} p_{ij} \log_{2}p_{ij} 
\end{equation}

\subsection{Segmentation}\label{segment_lab}
Each driving scenario can be imagined as a combination of multiple segments. These segments are either separated using road objects (e.g., intersections and stop signs) or road users (e.g., a slow lead vehicle). In order to analyze the driving behavior data, it is first required to find different segments of interest \cite{li2021extraction}. This is mostly because behaviors vary across segments, in which a person might be comfortable with different levels of acceleration/deceleration in a city segment, while the same acceleration/deceleration level may not feel suitable for a highway segment. We propose using Bayesian Change Point detection for driving behavior data segmentation in which different driving segments can be separated using change points in the distribution of vehicle's kinematic data (e.g., sudden acceleration) (Fig. \ref{fig:framework-analysis} - B). In this regard we first provide a brief overview on Bayesian Change Point detection method. A vector of multiple kinematic signals is then used as the input to the Bayesian Change Point detector.

\subsubsection{Bayesian Change Point Detection (BCP)}

Previous studies have pointed out the utility of BCP in detecting changes in the underlying distribution of data in different fields such as health \cite{malladi2013online}, epidemiology \cite{kass2012application}, ecology \cite{calder2017extensive}, and transportation engineering \cite{tavakoli2021harmony}. More specifically in transportation engineering, \cite{tavakoli2021harmony} used BCP to detect changes in driver's HR in naturalistic scenarios, which might be indicative of presence of stressors on the road. In another study Maye et al. \cite{maye2011bayesian} also used BCP to find different segments in vehicle's telemetry data. To perform BCP, we use the Bayesian change point model provided in Barry and Hartigan's book \cite{barry1993bayesian}. In summary this model assumes that the mean of the input (e.g., acceleration of the vehicle) within different segments remains constant. Below we formally define BCP for a univariate case. The reader is also referred to \cite{barry1993bayesian,maye2011bayesian,erdman2007bcp} for more information and details on this method.

Assuming we have $n$ input data point (e.g., vehicle's acceleration data points) $\{a_1,...,a_n\}$, we will use $a_{ij}$ to refer to the accelerometer data between indices $i$ and $j$. Let $\rho = (U_1,...,U_n)$ indicate a segment of the time series inside the non overlapping acceleration segments. A boolean array of change points is then used to to denote the segments. At each time step, if $U_i$ takes a value $1$, we have a new acceleration segment; else we remain in the same segment.

We are interested in the posterior density
$f(\rho|X)$. By Baye's theorem, this can be written as

\begin{equation}
f(\rho|X) \propto f(X|\rho) f(\rho)
\end{equation}

The Prior cohesion density can be retrieved as follows:
Let $p$ denote the probability of finding a new change point at each data point. We assume this probability to be the same at each data point. If we assume that there are $b$ segments, the prior cohesion density can be written as
\begin{equation}
f(\rho|p) = p^{b-1}(1-p)^{n-b}
\end{equation}

The joint density of observations and parameters given $\rho$ is a product of densities of different segments in $\rho$. Let us consider a single segment. If we assume that the data in this block is generated by a Gaussian process with mean $\theta$ and variance $\sigma^2$, let the prior density of $\theta$ be a Gaussian with mean $\mu_0$ and variance $\sigma_0^2$

\begin{equation}
\begin{split}
f(X_{ij}, \theta) = \Pi f(X_k|\theta) f(\theta)\\
f(X_{ij}) = \int \Pi f(X_k|\theta) f(\theta) d\theta
\end{split}
\end{equation}

We can then simplify the above integral to:

\begin{equation}
\begin{split}
f(X_{ij}) = (\frac{1}{2\pi \sigma^2})^{(j-i)/2} (\frac{\sigma^2}{\sigma_0^2 + \sigma^2})^{1/2} exp(V_{ij})
\end{split}
\end{equation}

Where
\begin{equation}
 V_{ij} = -\frac{\sum_{l=i+1}^{l=j} (X_l - \hat{X}_{ij})^2}{2\sigma^2} - \frac{(j-i) (\hat{X}_{ij} - \mu_0)^2}{2(\sigma^2 + \sigma_0^2)} 
\end{equation} 
and $\hat{X}_{ij}$ is the mean of the observations in the partition. However $f(X_{ij})$ still depends on the parameters $\mu_0, \sigma^2, \sigma_0^2$. Defining $w=\frac{\sigma^2}{\sigma_0^2 + \sigma^2}$
and choosing the following priors for the parameters:

\begin{equation}
\begin{split}
f(\mu_0) = 1 , -\infty \leq \mu_0 \le \infty\\
f(p) = 1/p_0, 0 \leq p \leq p_0 \\ 
f(\sigma^2) = 1/\sigma^2, 0\leq \sigma^2 \le \infty \\
f(w) = 1/w_0, 0 \leq w \leq w_0
\end{split}
\end{equation}
\begin{equation}
\begin{split}
 f(X|\rho, \mu_0, w) = \int_{0}^{\infty} 1/\sigma^2 \prod_{ij \epsilon P}^{} f(X_{ij}) d\sigma^2
 \end{split}
\end{equation}

After integrating out $\mu_0$ and $w$, This can be simplified to the indefinite integral below. The full derivation is provided in \cite{barry1993bayesian}.
\begin{equation}
f(X|\rho) \propto \int_{0}^{w_0} \frac{w^{(b-1)/2}}{(W + Bw)^{(n-1)/2}}dw,
\end{equation}

where
\begin{equation}
\begin{split}
\hat{X} = \sum_{i=1}^{n} X_i/n, \;  B = \sum_{ij \epsilon P} (j-i) (\hat{X}_{ij} - \hat{X})^2,  \\
 W = \sum_{ij \epsilon P} \sum_{l=i+1}^{l=j}(X_l - \hat{X}_{ij})^2
 \end{split}
\end{equation}

Similarly, after integrating out the change probability p, the prior cohesion density thus can be written as

\begin{equation}
f(\rho) \propto \int_{0}^{p_0} p^{b-1}(1-p)^{n-b} dp 
\end{equation}

To calculate the posterior distribution over segments, Markov Chain Monte Carlo (MCMC) is used \cite{gilks2005m}. A Markov chain is then defined with the following transition rule: with probability $p_i$, a new change point at the location $i$ is introduced. Here $B_1, W_1$ and $B_0, W_0$ refer to the expressions in (12) with and without the change point in location i.

\begin{equation}
\begin{split}
\frac{p_i}{1-p_i} &= \frac{p(U_i=1|X,U_j,j \neq i)}{p(U_i=0|X,U_j,j \neq i)}  \\
&= \frac{\int_{0}^{p_0} p^{b}(1-p)^{n-b-1} dp}{\int_{0}^{p_0} p^{b-1}(1-p)^{n-b} dp} x \frac{\int_{0}^{w_0} \frac{w^{b/2}}{(W_1 + B_1w)^{(n-1)/2}}dw}{\int_{0}^{w_0} \frac{w^{(b-1)/2}}{(W_0 + B_0w)^{(n-1)/2}}dw}
\end{split}
\end{equation}

Finally, this is simplified to a probabilistic model with the following two parameters $p_0$ and $w_0$. 

We use the package \textit{bcp} \cite{erdman2007bcp} in R programming language to implement our change point analysis. For the case study I, we use vehicle's kinematics as the main segmentation input. To this end, we use the input vector X for segmentation algorithm as combination of forward acceleration (a), lateral acceleration (l), and lateral rotational speed ($\omega_Z$):

$$X = \{a, l, \omega_Z\}$$

The vector X is then fed into BCP. The important point with respect to BCP is that it does not require to know the number of segments. This helps with longitudinal data where the actual number of driving segments is unknown. 

\begin{figure}
\begin{center}
  \includegraphics[width=0.95\linewidth]{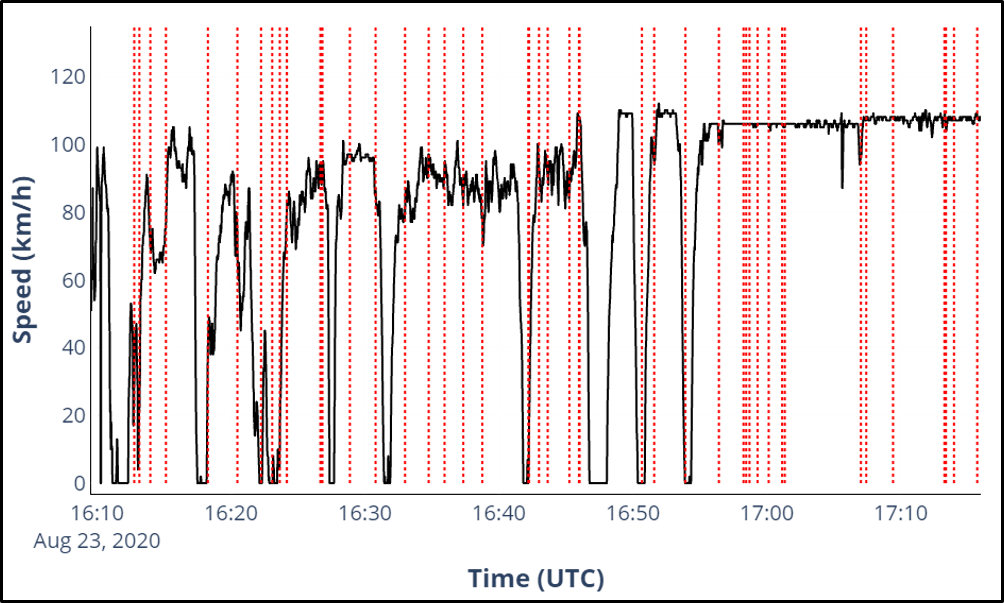}
  \end{center}
  \caption{Different segments of driving detected by the BCP applied on the kinematics data shown with dashed vertical lines.}
  \label{fig:segments}
\end{figure}

In order to illustrate the results of segmentation we use the vehicle's speed data. Note that the speed data was not used as the input into the segmentation as it is collected at 1 Hz, which is much lower than the IMU sensor (i.e., 100 Hz) and have not been used as the input to the segmentation. We use the speed data mostly due to the fact that speed can naturally show different segments of driving such as slowing down due to a lead vehicle, stopping for an intersection, and turning (Fig. \ref{fig:segments}). This helps with a better visual inspection of our results.  

For case study II, as we did not have access to an IMU sensor, we use vehicle's speed data, as well the acceleration, which is retrieved through applying a derivative on the speed data. The segmentation is performed separately for each participant. 

\subsection{Pattern Inference}\label{lab:pattern_inf}

In order to perform pattern inference we use a combination of Gaussian Mixture Modeling (GMM) and Latent Dirichlet Allocation (LDA), referred to as GMM-LDA \cite{agamennoni2014automated,agamennoni2014bayesian,prabhudesai2018augmented,li2021extraction}. LDA is a text analytics probabilistic hierarchical Bayesian model, which is used to infer topics from a set of words \cite{blei2003latent}. When applying it to driving data, the goal is to provide a main driving pattern and state pattern for each segment of driving. To do so, we first need to generate 50 word-like objects (i.e., discretize the data) from the continuous raw vehicle telemetry, gaze, and HR data using GMM (Fig. \ref{fig:framework-analysis} - C). GMM-LDA method then uses LDA on top of GMM objects to produce topics for each segment of driving, which can be considered a driving pattern (Fig. \ref{fig:framework-analysis} - D). Below we formally outline this method.

\subsubsection{GMM-LDA}
In LDA, each word is modeled as a finite random mixtures over an underlying set of k topics and each topic is represented as a distribution over a set of words \cite{blei2003latent}. A word (w) is defined as a basic unit of discrete data which is drawn from a set of vocabulary. In this way, a document is then a sequence of N words $N = \{w_1,w_2,...,w_n\}$ and a corpus is a set a of M documents \cite{blei2003latent}. LDA is based on the following generative assumptions \cite{blei2003latent,li2021extraction}:

\begin{equation}
    N \sim Poisson(\xi)
\end{equation}
\begin{equation}
    \theta_{m} \sim Dirichlet(\alpha)
\end{equation}
\begin{equation}
    \varphi_{k} \sim Dirichlet(\beta)
\end{equation}

which $\theta_{m}$ is the topic distribution of document m in M and  $\varphi_{k}$ is the distribution of each word w in topic k in K. Note that $\alpha$ and $\beta$ are the prior distributions of $\theta$ and $\varphi$ respectively. In this way, the topic of the nth word in document m is $Z_{m,n}$, which has a Multinomial distribution on $\theta_{m}$. Additionally, the nth word in document m has a Multinomial distribution over $\varphi_{z_{m,n}}$. 

Historically LDA is designed for inferring topics from a set of documents and is based on text type data, which is by nature a discrete type data. When considering continuous data sources (e.g., vehicle's acceleration), multiple methods were proposed to overcome this issue. A more common approach is to apply GMM clustering method on the continuous data to generate word-like objects \cite{agamennoni2014automated,agamennoni2014bayesian,prabhudesai2018augmented,li2021extraction}. GMM is a clustering method that identifies a mixture of multidimensional Gaussian probability distributions that can best describe the input vector \cite{reynolds2009gaussian}. Formally, a GMM model is a weighted summation of M different components as:

\begin{equation}
    p(x|\lambda) = \sum_{i=1}^{M} w_i g(x|\mu_i,\Sigma_i)
\end{equation}

In the equation above, x is the multidimensional vector and $w_i$ is the identified different Gaussian distributions that sum up to 1, and are described as:
\begin{equation}
    g(x|\mu_i,\Sigma_i) = \frac{1}{(2\pi)^{\frac{D}{2}}|\Sigma_i|^{\frac{1}{2}}} \exp{(-\frac{1}{2}(x-\mu_i)'\Sigma_i^{-1}(x-\mu_i))}
\end{equation}

in which $\mu_i$ is the mean vector and $\Sigma_i$ is the covariance matrix. The parameters of different Gaussian distributions can be retrieved through the Expectation Maximization algorithm which is described in detail in \cite{reynolds2009gaussian}.

We first apply GMM algorithm to obtain word-like objects (i.e., descretizing data) from the multimodal vehicle kinematic data, which includes forward acceleration (a), lateral acceleration (l), and lateral rotational speed ($\omega_Z$) (Fig. \ref{fig:framework-analysis} - C). These inputs are the same inputs that were used  in  segmentation section (\ref{segment_lab}). Based on the previous literature we chose 50 as the number of words \cite{li2021extraction}. Also, note that using Bayesian Information Criterion (BIC) for defining the number of components, we observed very small enhancements ($<0.0001)$ in the BIC value after increasing the number of words above 50.

\section{Results}
\begin{figure*}[h]
\begin{center}
  \includegraphics[width=\linewidth]{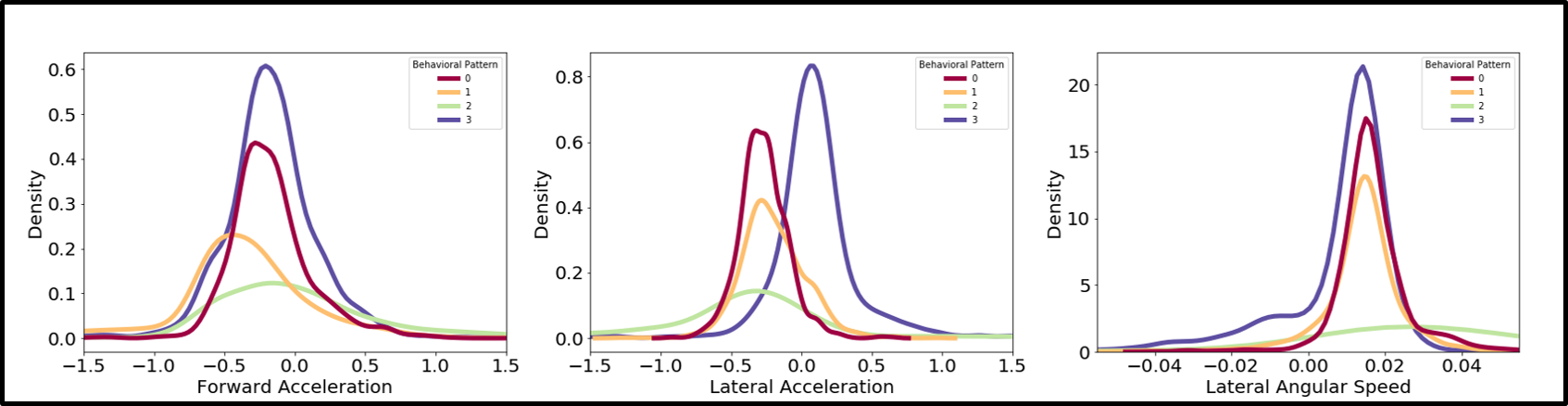}
  \end{center}
  \caption{Distribution of different kinematic sensor readings within different driving patterns. Note that differences across patterns in the kinematic sensor distributions, for instance, pattern 3 has a more positive lateral acceleration than the other 3 patterns. Also note that for some patterns the difference is only visible through one sensor such as the comparison between patterns 0 and 3 lateral acceleration.}
  \label{fig:dist_cluster}
\end{figure*}
Based on the generated GMM distribution, we then apply LDA to find the driving pattern that best describes each segment in the driving scenario (Fig. \ref{fig:framework-analysis} - E). When using LDA, we need to define the number of topics prior to analysis. Driving scenarios often consist of different patterns \cite{li2021extraction}. For instance, braking behavior includes a limited number of patterns which can be high negative acceleration (i.e., harsh brake), or a low negative acceleration (i.e., normal brake). Thus it is likely that there are more than one driving pattern present in an individual's driving data \cite{li2021extraction}. While previous studies suggested the number of patterns to be 4 we also tested different numbers ranging from 2-6 \cite{bender2015unsupervised,li2021extraction}. We observed that with increasing the number of topics in LDA (patterns in driving) more than 4, some of the topics become very similar from the visual inspection of the distributions so it becomes impractical to have them as separate patterns. For instance, moving from 4 to 5 topics, two of the generated topics will be similar in their mean and standard deviation of acceleration, and speed. Additionally, reducing the topics from 4 to 3 misses the difference in the braking patterns.

For the case study II, we also performed segmentation using vehicle's speed and acceleration. Then for each segment we followed the GMM-LDA method and applied it to the driving behavior as well as driver state data to find patterns of behaviors and state in each segment. Below we first present the results for case study I, which helps better illustrate how the method is applied. We then discuss the results of applying the same method to a larger pool of participants.

In this section we present the results of applying the GMM-LDA method with Bayesian Change Point segmenting on two separate case studies. As described in section \ref{sec:case_study_data}, the first case study is from one participant with high resolution kinematic data, while the second case study is from 12 participants vehicles that had low resolution kinematic data.

\section{Results - Case Study I}
In this section, we showcase how GMM-LDA method can be applied on driver behavior and state data to find patterns in driver's behavior (e.g., harsh braking) and states (e.g., high stress level). We first discuss the patterns detected in vehicle kinematic data. We then move to driver's state data (e.g., HR). Fig. \ref{fig:dist_cluster} shows the distribution of kinematic data across the four different detected driving patterns (refer to section \ref{lab:pattern_inf} for details on pattern recognition). As shown on this figure, each behavior pattern has its own distinct distribution. For instance, when comparing patterns 3 and 2, we observe the difference in the location of peak in the forward acceleration as well as lateral acceleration. However, comparing across patterns 0 and 3, we only observe a major difference when comparing the distribution of their lateral acceleration.

Additionally, Table \ref{tab:detail_of_behaviors} provides the detailed statistics of each driving pattern. The first step is to confirm that the driving patterns detected by GMM-LDA are as a result of different underlying distributions and are statistically different. The Kolmogorov-Smirnov test between different distributions is performed \cite{massey1951kolmogorov}. The pairwise comparison of all distributions were significant at $p<0.05$ level and is shown on Table \ref{tab:stats_behavior}. Below we elaborate in detail on each driving pattern detected in the kinematic data and how they relate to certain predefined driving behaviors.

\begin{table}[ht]
\caption{Details of statistical tests on different patterns detected through unsupervised modeling}
\label{tab:stats_behavior}
\resizebox{0.48\textwidth}{!}{%
\begin{tabular}{ccccc}
Source                                                                           & Comparison & KS Test Statistic & p-value           & Significant at 0.05? \\ \hline
\multirow{6}{*}{\begin{tabular}[c]{@{}c@{}}Forward \\ Acceleration\end{tabular}} & 0\&1       & 0.3369            & \textless{}0.0001 & Y                    \\ \cline{2-5} 
                                                                                 & 0\&2       & 0.2467            & \textless{}0.0001 & Y                    \\ \cline{2-5} 
                                                                                 & 0\&3       & 0.0622            & 0.0003            & Y                    \\ \cline{2-5} 
                                                                                 & 1\&2       & 0.3447            & \textless{}0.0001 & Y                    \\ \cline{2-5} 
                                                                                 & 1\&3       & 0.3021            & \textless{}0.0001 & Y                    \\ \cline{2-5} 
                                                                                 & 2\&3       & 0.2133            & \textless{}0.0001 & Y                    \\ \hline
\multirow{6}{*}{\begin{tabular}[c]{@{}c@{}}Lateral \\ Acceleration\end{tabular}} & 0\&1       & 0.1670            & \textless{}0.0001 & Y                    \\ \cline{2-5} 
                                                                                 & 0\&2       & 0.2761            & \textless{}0.0001 & Y                    \\ \cline{2-5} 
                                                                                 & 0\&3       & 0.7096            & \textless{}0.0001 & Y                    \\ \cline{2-5} 
                                                                                 & 1\&2       & 0.2883            & \textless{}0.0001 & Y                    \\ \cline{2-5} 
                                                                                 & 1\&3       & 0.5568            & \textless{}0.0001 & Y                    \\ \cline{2-5} 
                                                                                 & 2\&3       & 0.6054            & \textless{}0.0001 & Y                    \\ \hline
\multirow{6}{*}{\begin{tabular}[c]{@{}c@{}}Angular \\ Velocity\end{tabular}}     & 0\&1       & 0.1201            & 0.0003            & Y                    \\ \cline{2-5} 
                                                                                 & 0\&2       & 0.4365            & \textless{}0.0001 & Y                    \\ \cline{2-5} 
                                                                                 & 0\&3       & 0.2642            & \textless{}0.0001 & Y                    \\ \cline{2-5} 
                                                                                 & 1\&2       & 0.4520            & \textless{}0.0001 & Y                    \\ \cline{2-5} 
                                                                                 & 1\&3       & 0.1488            & \textless{}0.0001 & Y                    \\ \cline{2-5} 
                                                                                 & 2\&3       & 0.5152            & \textless{}0.0001 & Y                   
\end{tabular}%
}
\end{table}

\begin{figure*}
\begin{center}
  \includegraphics[width=\linewidth]{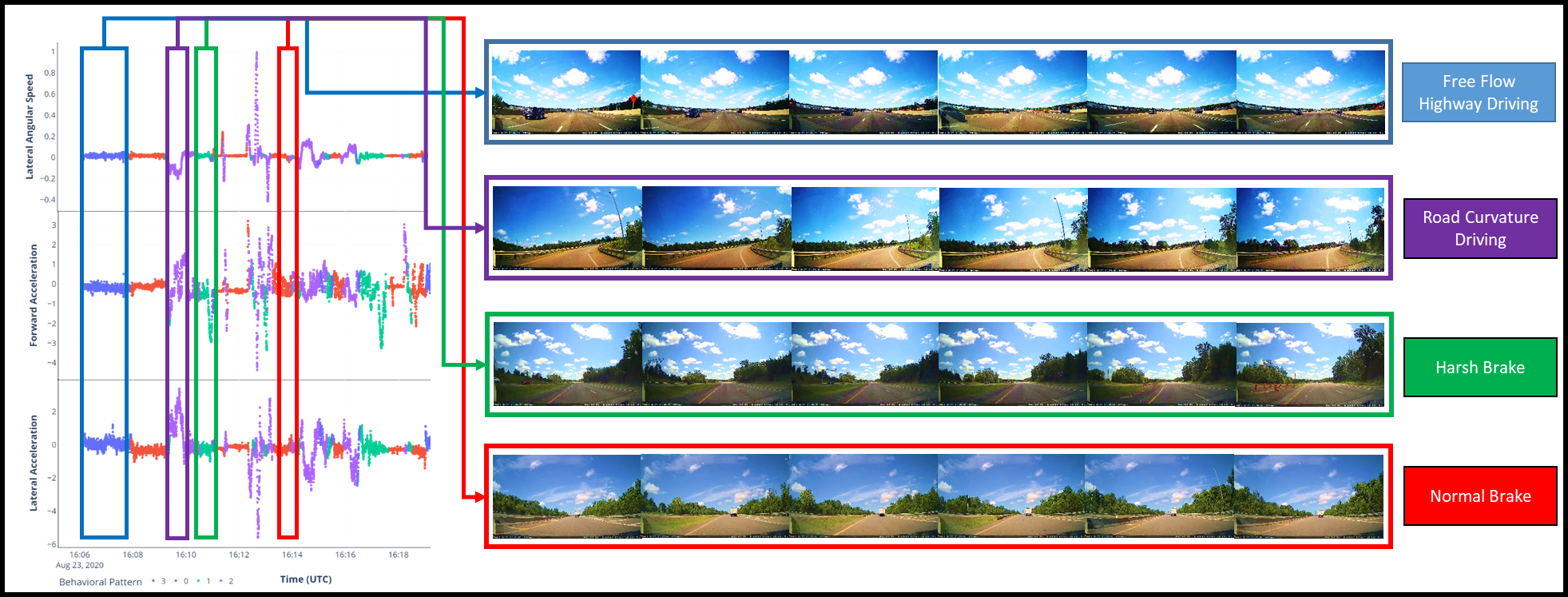}
  \end{center}
  \caption{Samples of the detected driving patterns. Patterns 0 and 1 are associated with normal and harsh braking behavior, while pattern 2 is associated to curve driving behavior, and pattern 3 is related to highway free flow driving behavior.}
  \label{fig:sample_pattern}
\end{figure*}

Considering the differences in sensor readings across the driving patterns, we find dominant differences that can be associated to certain well-known driving behaviors. Driving patterns 0 and 1 both exhibit high negative forward acceleration (deceleration) values (-0.1667 and -0.4180) respectively. These two driving patterns are mostly related to braking behavior. The more negative value of mean forward acceleration in driving pattern 1 suggest a harsher braking pattern. Note in Table \ref{tab:detail_of_behaviors} the mean speed of the two driving patterns are similar at 76.27 and 76.61 km/h respectively. This provides evidence for the two patterns to be similar in nature, as both are related to braking. We refer to pattern 0 as normal braking behavior, and pattern 1 as harsh braking behavior. Driving pattern 2 has a relatively flat forward acceleration distribution including a broad range of forward accelerations with a mean value 0.032 which is close to zero. Due to exhibiting an average higher rotational velocity (0.027), and a broader lateral acceleration distribution, this driving pattern might be associated to normal road curvatures. We refer to this driving pattern as road curvature driving behavior. Driving pattern 3 has the highest mean speed value (89 km/h), with a positive forward acceleration (0.095). This driving pattern also exhibits one large peak at a positive lateral angular speed at 0.0087 which is very close to zero indicating straight driving. This pattern can be associated to normal free flow driving in the highway, and we refer to it as the highway free flow driving behavior. A sample of such behaviors are also shown on Fig. \ref{fig:sample_pattern}, where snapshots of different behaviors (e.g., free flow highway driving) are shown from the dataset.

\begin{table}[ht]
\caption{Detailed statistics of forward acceleration, lateral acceleration, angular velocity, and speed of each recognized behavior. Note that different patterns resemble well-known driving behaviors.}
\label{tab:detail_of_behaviors}
\resizebox{0.48\textwidth}{!}{%
\begin{tabular}{llllll}
\hline
Driving Behavior Data &
  Statistical Index &
  \multicolumn{1}{c}{\begin{tabular}[c]{@{}c@{}}Driving Pattern 0\\ (Normal Braking)\end{tabular}} &
  \multicolumn{1}{c}{\begin{tabular}[c]{@{}c@{}}Driving Pattern 1\\ (Harsh Braking)\end{tabular}} &
  \multicolumn{1}{c}{\begin{tabular}[c]{@{}c@{}}Driving Pattern 2\\ (Road Curvature \\ Driving)\end{tabular}} &
  \multicolumn{1}{c}{\begin{tabular}[c]{@{}c@{}}Driving Pattern 3\\ (Free Flow Driving)\end{tabular}} \\ \hline
\multirow{5}{*}{Forward Acceleration} & Mean               & -0.16677 & -0.41807 & 0.032059 & -0.17298 \\
                                      & Standard Deviation & 0.343867 & 0.617844 & 0.677998 & 0.397301 \\
                                      & 25th percentile    & -0.34826 & -0.61339 & -0.41371 & -0.3527  \\
                                      & Median             & -0.20286 & -0.40094 & -0.08027 & -0.18616 \\
                                      & 75th percentile    & -0.02847 & -0.13217 & 0.303769 & 0.0067   \\ \hline
\multirow{5}{*}{Lateral Acceleration} & Mean               & -0.26936 & -0.20002 & -0.35933 & 0.095032 \\
                                      & Standard Deviation & 0.205908 & 0.263469 & 0.774748 & 0.347671 \\
                                      & 25th percentile    & -0.38046 & -0.36103 & -0.66054 & -0.0605  \\
                                      & Median             & -0.27365 & -0.21968 & -0.345   & 0.077288 \\
                                      & 75th percentile    & -0.15625 & -0.04575 & -0.08588 & 0.221336 \\ \hline
\multirow{5}{*}{Rotational velocity}  & Mean               & 0.016956 & 0.01432  & 0.027152 & 0.008777 \\
                                      & Standard Deviation & 0.012567 & 0.012698 & 0.073349 & 0.018666 \\
                                      & 25th percentile    & 0.011648 & 0.00877  & 0.012364 & 0.004099 \\
                                      & Median             & 0.015752 & 0.014653 & 0.02795  & 0.012516 \\
                                      & 75th percentile    & 0.021457 & 0.020216 & 0.047836 & 0.01734  \\ \hline
\multirow{5}{*}{Speed}                & Mean               & 76.27409 & 76.61246 & 79.16773 & 89.84004 \\
                                      & Standard Deviation & 36.47230  & 29.24584 & 27.87107 & 27.15401 \\
                                      & 25th percentile    & 65       & 53       & 67       & 90       \\
                                      & Median             & 97       & 88       & 89       & 98       \\
                                      & 75th percentile    & 98       & 97       & 99       & 106      \\ \hline
\end{tabular}%
}
\end{table}

\begin{figure}
\begin{center}
  \includegraphics[width=\linewidth]{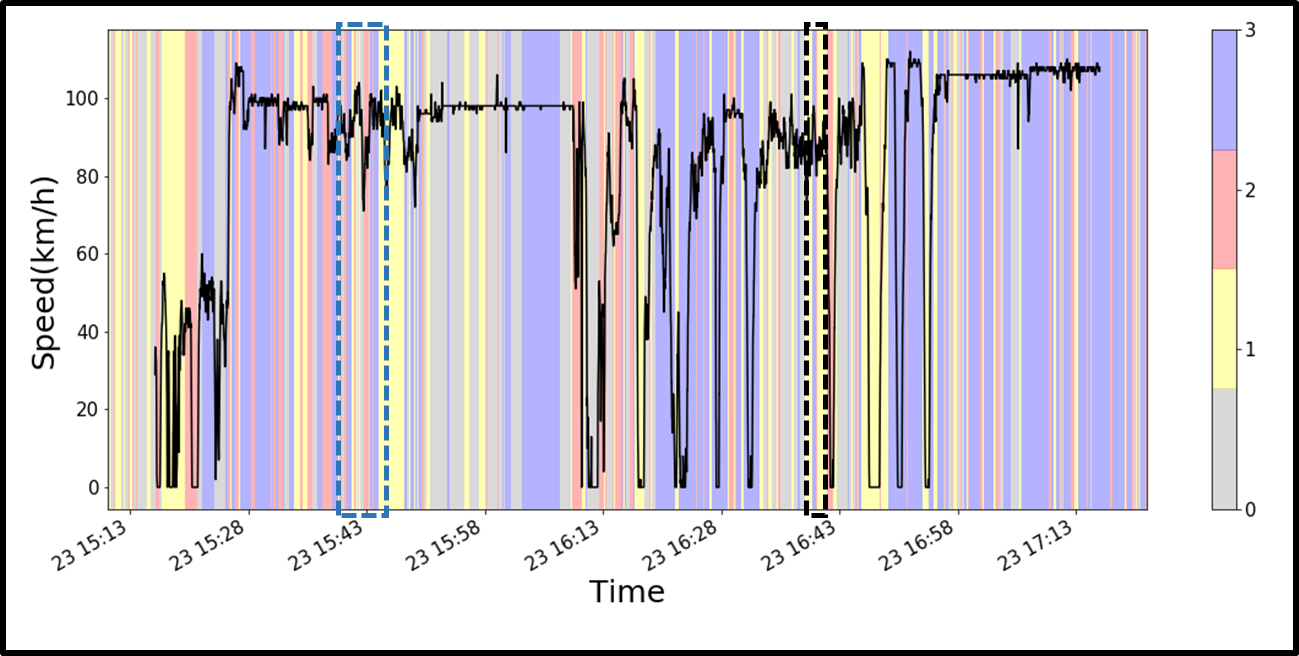}
  \end{center}
  \caption{The change in driving patterns throughout the different segments of driving. Each segment is associated with one of the four driving patterns as described previously. Note that when speed changes more sporadically, the behavior also changes more often (e.g., blue dashed box). Also, the black dashed box shows a sequence of sample behavior starting from free flow highway driving, switching to road curvature driving, and finishing with a harsh brake.}
  \label{fig:driving_cluster_seg}
\end{figure}

Additionally, the sequence of different detected patterns in kinematic sensors can provide insights into driver's behavior. We discuss the findings when analyzing the sequence of behaviors through a case study and a transition matrix of behavior sequences. First through a case study, Fig. \ref{fig:driving_cluster_seg} depicts the sequence of driving behaviors through parts of the driving scenario. As visually observed, in locations that speed varies sporadically, the diver behavior also switches more often (e.g., 15:43 until 15:46 shown with a blue dashed box on Fig. \ref{fig:driving_cluster_seg}). To quantify these sequences, as shown in Table \ref{tab:transition_matrix}, the probability of transition between the different driving behaviors as recognized by our method. Each element of the table at location i, and j shows the $p_{i,j}$ transition probability between the elements i and j. As shown on the table, when in normal braking behavior, the chances of switching to behaviors of road curvature driving behavior and free flow driving is higher than harsh braking behavior. However, when being at harsh braking behavior, the driver is most likely to switch to normal braking than other two behaviors. Looking at the free flow driving behavior, the driver is most likely to continue with the high speed highway driving ($p_{3,3} = 0.55$), and the probability of switching to a harsh braking behavior ($p_{3,1} = 0.13$) is equal to the normal braking behavior, while being less than switching to the curve driving behavior. A sample of a probable sequence can be to start from free flow highway driving, switch to road curvature driving and attempt a harsh brake. For a better illustration, such a sample sequence is shown with black dashed box on Fig. \ref{fig:driving_cluster_seg}.

\begin{table}[]
\caption{Transition matrix between different driving behaviors for case study I dataset.}
\label{tab:transition_matrix}
\centering
\resizebox{0.40\textwidth}{!}{%
\begin{tabular}{l|llll}
\textbf{Driving Behavior} &
  \textbf{\begin{tabular}[c]{@{}l@{}}Normal \\ Braking\end{tabular}} &
  \textbf{\begin{tabular}[c]{@{}l@{}}Harsh\\ Braking\end{tabular}} &
  \textbf{\begin{tabular}[c]{@{}l@{}}Road Curvature\\ Driving\end{tabular}} &
  \textbf{\begin{tabular}[c]{@{}l@{}}Free Flow\\ Driving\end{tabular}} \\ \hline
\textbf{\begin{tabular}[c]{@{}l@{}}Normal \\ Braking\end{tabular}}  & 0.28 & 0.21 & 0.26 & 0.26 \\
\textbf{\begin{tabular}[c]{@{}l@{}}Harsh\\ Braking\end{tabular}}    & 0.27 & 0.27 & 0.21 & 0.25 \\
\textbf{\begin{tabular}[c]{@{}l@{}}Road Curvature\\ Driving\end{tabular}}     & 0.25 & 0.27 & 0.26 & 0.21 \\
\textbf{\begin{tabular}[c]{@{}l@{}}Free Flow\\ Driving\end{tabular}} & 0.13 & 0.13 & 0.19 & 0.55
\end{tabular}%
}
\end{table}

\begin{figure}
\begin{center}
  \includegraphics[width=\linewidth]{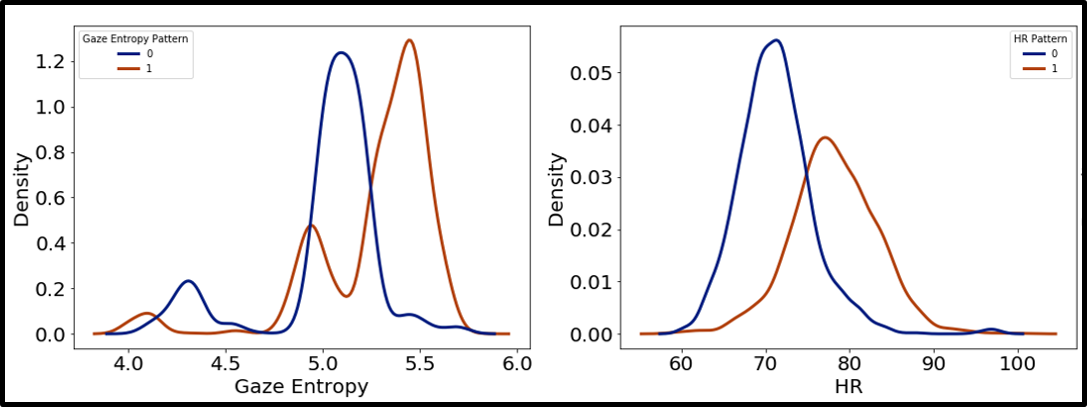}
  \end{center}
  \caption{Distribution of different driver state patterns detected through HR and gaze entropy. Note the differences across the distributions pointing to different states for the driver}
  \label{fig:dist_state_hr_gaze}
\end{figure}

The GMM-LDA method was then applied to the driver's gaze and HR data. We considered two patterns for these two modalities relating to normal and abnormal (high) for HR, and low and high for gaze entropy, respectively. Fig. \ref{fig:dist_state_hr_gaze} shows the different driver state patterns detected through driver's gaze entropy and HR data. We first confirmed that the distribution of normal/abnormal HR and low/high gaze are statistically significant. Similar to previous sections this is performed through Kolmogorov-Smirnov test \cite{massey1951kolmogorov} and the results of comparison between different distribution where significant at $p<0.0001$ level. Table \ref{tab:statsstates} shows the significance of the tests across the different distributions.

Driver state pattern 0 in HR is related to normal HR values, which has a mean value of 71.23 bpm with an standard deviation of 4.47 bpm. We refer to this driver state pattern as the normal HR or calm state. On the other hand, driver state pattern 1 is related to high HR with a mean of 78 bpm and a standard deviation of 5.13 bpm, which we refer to as abnormal HR depicting a situation of having high stress driver state. As shown on Fig. \ref{fig:dist_state_hr_gaze}, density of the abnormal HR is lower, indicating the less amount of time spent in this state pattern throughout all observations. This is also in line with our previous study \cite{tavakoli2021harmony} showing that driver's HR through stressors on the road increases from it's baseline values for a short period of time and moves back to it's baseline. 

\begin{table}[]
\caption{Details of statistical tests across different distributions of HR and gaze data}
\label{tab:statsstates}
\resizebox{0.47\textwidth}{!}{%
\begin{tabular}{ccccc}
Driver State Data & Comparison               & KS Test Statistic & p-value           & Significant at 0.05? \\ \hline
HR     & Abnormal \& Normal HR    & 0.5999            & \textless{}0.0001 & Y                    \\ \hline
Gaze   & Low \& High Gaze Entropy & 0.6688            & \textless{}0.0001 & Y                    \\ \hline
\end{tabular}%
}
\end{table}

\begin{table}[]
\caption{Detailed statistics of HR, and gaze for each recognized state pattern}
\label{tab:state_cluster_details}
\resizebox{0.48\textwidth}{!}{%
\begin{tabular}{llll}
\hline
Driver State Data &
  Statistical Index &
  \begin{tabular}[c]{@{}l@{}}State Pattern 0\\ (Normal HR/\\ Low Gaze Entropy)\end{tabular} &
  \begin{tabular}[c]{@{}l@{}}State Pattern 1\\ (Abnormal HR/\\ High Gaze Entropy)\end{tabular} \\ \hline
\multirow{5}{*}{HR} & Mean               & 71.23 & 78.13 \\ 
                            & Standard Deviation & 4.47  & 5.13  \\ 
                            & 25th percentile    & 68.4    & 75  \\ 
                            & Median             & 71    & 78    \\ 
                            & 75th percentile    & 73.7  & 81.4  \\ \hline 
\multirow{5}{*}{Gaze}       & Mean               & 5.01  & 5.27  \\
                            & Standard Deviation & 0.30  & 0.31  \\
                            & 25th percentile    & 4.99  & 5.12  \\
                            & Median             & 5.08  & 5.36  \\
                            & 75th percentile    & 5.17  & 5.47  \\ \hline
\end{tabular}%
}
\end{table}

\begin{figure}
\begin{center}
  \includegraphics[width=\linewidth]{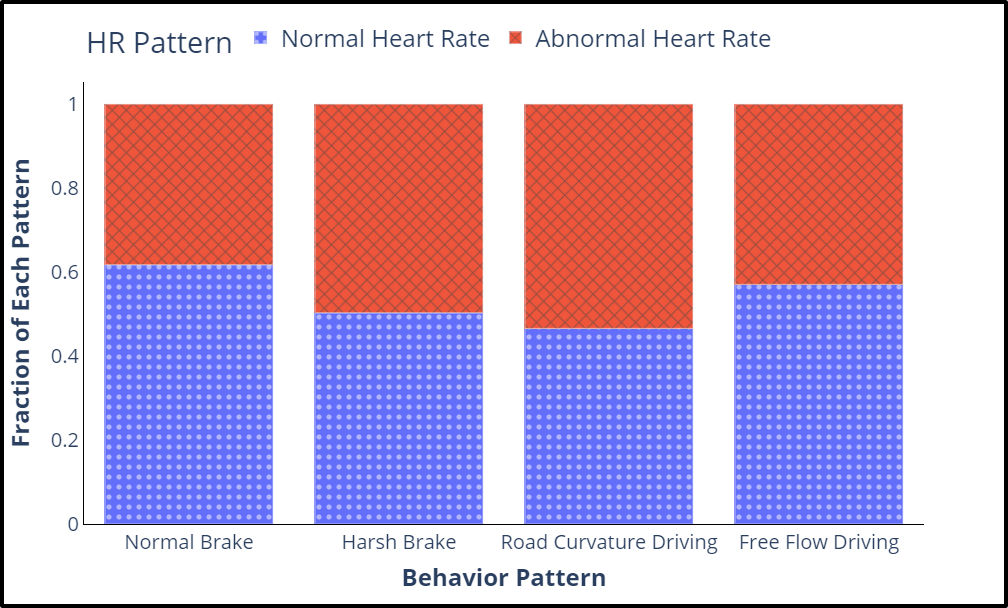}
  \end{center}
  \caption{The co-occurence of driving clusters with clusters of HR}
  \label{fig:driving_cluster_HR}
\end{figure}

Fig. \ref{fig:driving_cluster_HR} shows the fraction of normal and abnormal HR in the recognized driving patterns such as normal brake, and harsh brake. As shown, the driver state pattern with abnormal HR happens more often than the normal HR in the road curvature driving behavior. A Chi-Square test \cite{mchugh2013chi} also shows that the ratio between the counts of each state is significantly different than being equal (i.e., ratio of 1) with a $\chi^2 = 43.87$ and a $p-value < 0.0001$.  This figure also shows that the driver has more abnormal HR states during harsh brake compared to normal brakes, in which a Kruskal-Wallis test \cite{kruskal1952use} confirms the figure with a $statistic = 404.57$ and a $p-value<0.0001$. Considering that normal HR indicates a calmer driving state, this implies the driver is calmer during normal braking behavior compared to harsh braking behaviors. Additionally, the driver had a higher level of normal HR as compared to abnormal HR in free flow driving behavior, in which we confirmed this through a Chi-Square test with a $\chi^2 = 558.6$ and a $p-value < 0.0001$. 

Driver state pattern 0 in gaze with a mean of 5.01, is related to samples of gaze with lower GTE. On the other hand, driver state pattern 1 has a mean value of 5.27. As mentioned in section \ref{sec:background}, higher GTE in pattern 1 might be associated to higher task demand, higher scene complexity, and work load (Table \ref{tab:state_cluster_details}).

\begin{figure}
\begin{center}
  \includegraphics[width=\linewidth]{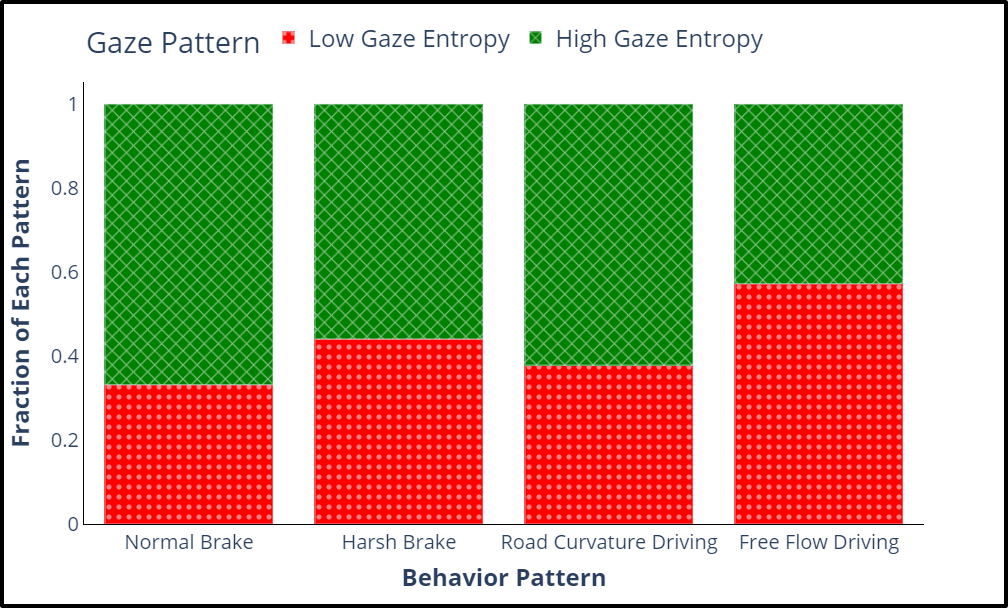}
  \end{center}
  \caption{The co-occurence of driving clusters with clusters of gaze entropy}
  \label{fig:driving_cluster_gaze}
\end{figure}

Fig. \ref{fig:driving_cluster_gaze} shows the fraction of low and high GTE in the recognized driving patterns such as normal brake, and harsh brake. As shown, in the free flow driving, low GTE pattern happens more often than high GTE which might indicate lower work load for this driver during free flow driving. A Chi-Square test \cite{mchugh2013chi} shows the counts of each low and high GTE within free flow driving is significantly different than equal probability (i.e., ratio of 1) with a $\chi^2 = 587.26$ and a $p-value < 0.0001$. Also, the ratio of high to low GTE in normal braking is higher than that of harsh braking, which implies that the probability of having a higher workload in normal braking is more than harsh braking. This is confirmed with A Kruskal-Wallis test \cite{kruskal1952use} indicating a $statistic = 88.31$, and a $p-value<0.0001$. This can be due to the fact that in normal braking the period of time that the driver is in the process of braking is higher, thus driver's GTE and the associated workload are more likely to be higher.

\section{Results - Case Study II}
\begin{figure*}
\begin{center}
  \includegraphics[width=\linewidth]{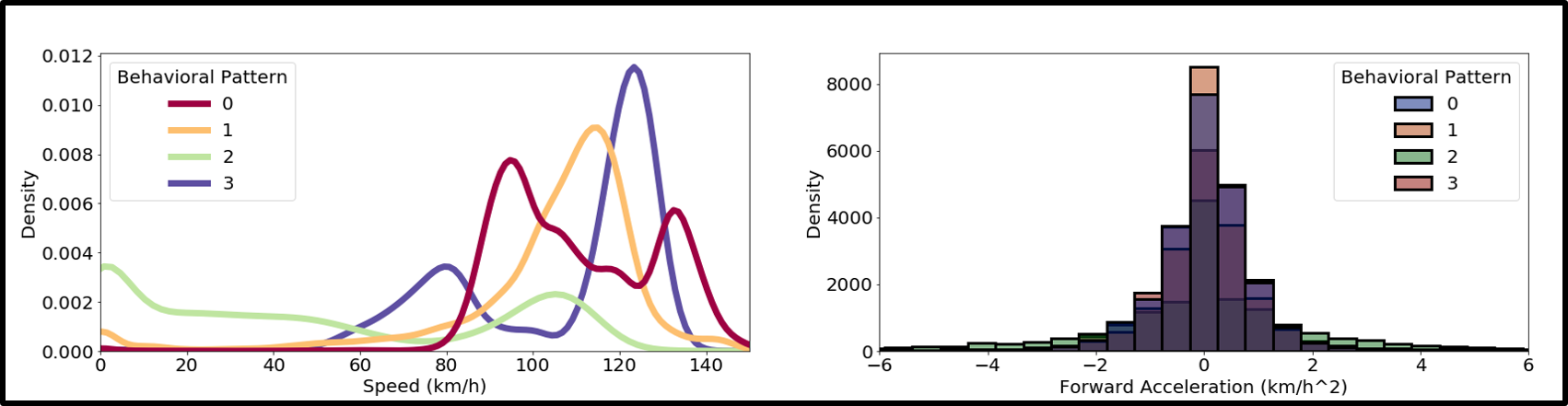}
  \end{center}
  \caption{The distribution of forward acceleration and speed across different patterns retrieved unsupervised from 12 participants. Note that because we retrieved accelerometer data from the low resolution speed, the distributions are less smooth as compared to case study I. Thus, we are visualizing the differences across driving patterns with a histogram instead of a kernel density estimation (KDE).}
  \label{fig:acc_cluster_case_II}
\end{figure*}

Similar to the case study I, we performed segmentation, clustering and pattern inference on the case study II dataset. Note that case study II includes kinematic data from vehicle's speed and acceleration from 12 participants. Fig. \ref{fig:acc_cluster_case_II} depicts the distribution of speed and accelerometer data across the different detected driving patterns. Similar to case study I, each of the different detected patterns exhibit distinct characteristics. For instance, comparing across patterns 0, and 1, we observe the difference in the location of the peak in the distribution. In order to confirm that the patterns are statistically different, we have applied the Kolmogorov-Smirnov test \cite{massey1951kolmogorov}. The results of the pairwise tests across patterns are depicted in Table \ref{tab:test_behavior_case_II} and are all statistically significant.

\begin{table}[]
\caption{Details of statistical tests performed on the detected driving patterns. The tests shows that the distributions of speed and accelerometer data are significantly different across the detected patterns.}
\label{tab:test_behavior_case_II}
\resizebox{0.48\textwidth}{!}{%
\begin{tabular}{ccccc}
Source                                                                           & Comparison & KS Test Statistic & p-value           & Significant at 0.05? \\ \hline
\multirow{6}{*}{\begin{tabular}[c]{@{}c@{}}Forward \\ Acceleration\end{tabular}} & 0\&1       & 0.04              & \textless{}0.0001 & Y                    \\ \cline{2-5} 
 & 0\&2 & 0.08 & \textless{}0.0001 & Y \\ \cline{2-5} 
 & 0\&3 & 0.08 & \textless{}0.0001 & Y \\ \cline{2-5} 
 & 1\&2 & 0.08 & 0.0003            & Y \\ \cline{2-5} 
 & 1\&3 & 0.09 & \textless{}0.0001 & Y \\ \cline{2-5} 
 & 2\&3 & 0.02 & \textless{}0.0001 & Y \\ \hline
\multirow{6}{*}{Speed}                                                           & 0\&1       & 0.7               & \textless{}0.0001 & Y                    \\ \cline{2-5} 
 & 0\&2 & 0.9  & \textless{}0.0001 & Y \\ \cline{2-5} 
 & 0\&3 & 0.9  & \textless{}0.0001 & Y \\ \cline{2-5} 
 & 1\&2 & 0.9  & \textless{}0.0001 & Y \\ \cline{2-5} 
 & 1\&3 & 0.8  & \textless{}0.0001 & Y \\ \cline{2-5} 
 & 2\&3 & 0.5  & \textless{}0.0001 & Y \\ \hline
\end{tabular}%
}
\end{table}

Considering the differences in vehicle's speed and acceleration across the driving patterns, we find distinct differences associated to certain well-known driving behaviors, such as differences in mean and standard deviation as depicted in Table \ref{tab:stat_behavior_case_II}. Similar to case study I, driving pattern 0 is related to normal brakes, with a negative acceleration (-0.006). Additionally, driving pattern 1 is related to harsh brakes with the highest negative acceleration and the lowest mean speed as compared to all the other patterns. The high value of the standard deviation of speed (42.01) in the driving pattern 1 confirms that this behavior can happen at different speeds depending on the environment, and it can quickly change the speed from high to lower values. Driving pattern 2 is related to accelerating behavior, which has the highest positive acceleration (0.041). Similar to case study 1, driving pattern 3 exhibits a very low negative acceleration (-0.001) with the highest mean speed (110.368) and lowest standard deviation (18.78). This in fact, depicts the behavior through free flow driving where the driver keeps a constant speed while not accelerating/decelerating. Also, note the difference between driving pattern 0 and 3. Driving pattern 0 exhibits a higher standard deviation in speed relative to pattern 1 (23.4 vs. 18.74) with a lower mean speed value (107.195 vs. 110.368), respectively. These differences align with the difference between braking and highway driving behaviors. Please note that in case study II, we did not have access to lateral kinematic data thus we could not confirm the curve driving behavior across participants. 
\begin{table}[]
\caption{Detailed statistics of forward acceleration and speed in each recognized driving pattern. Note that each pattern exhibits certain characteristics that resembles a well-known driving behavior.}
\label{tab:stat_behavior_case_II}
\resizebox{0.49\textwidth}{!}{%
\begin{tabular}{clcccc}
\hline
\multicolumn{1}{l}{Behavioral Attribute} &
  Statistical Index &
  \begin{tabular}[c]{@{}c@{}}Driving Pattern 0\\ (Normal Braking)\end{tabular} &
  \begin{tabular}[c]{@{}c@{}}Driving Pattern 1\\ (Harsh Braking)\end{tabular} &
  \begin{tabular}[c]{@{}c@{}}Driving Pattern 2\\ (Accelerating)\end{tabular} &
  \begin{tabular}[c]{@{}c@{}}Driving Pattern 3\\ (Free Flow Driving)\end{tabular} \\ \hline
\multirow{5}{*}{\begin{tabular}[c]{@{}c@{}}Forward Acceleration\\ (km/h\textasciicircum{}2)\end{tabular}} & Mean               & -0.006  & -0.045  & 0.041   & -0.001  \\
                                                                                                          & Standard Deviation & 0.805   & 1.259   & 1.291   & 0.523   \\
                                                                                                          & 25th percentile    & -0.118  & -0.297  & -0.259  & -0.162  \\
                                                                                                          & Median             & 0.0001  & -0.0006 & 0.0038  & -0.0006 \\
                                                                                                          & 75th percentile    & 0.108   & 0.233   & 0.353   & 0.126   \\ \hline
\multirow{5}{*}{\begin{tabular}[c]{@{}c@{}}Speed\\ (km/h)\end{tabular}}                                   & Mean               & 107.195 & 47.872  & 103.252 & 110.368 \\
                                                                                                          & Standard Deviation & 23.404  & 42.014  & 25.964  & 18.784  \\
                                                                                                          & 25th percentile    & 84      & 6       & 99      & 96      \\
                                                                                                          & Median             & 119     & 38      & 110     & 107     \\
                                                                                                          & 75th percentile    & 124     & 96      & 117     & 128     \\ \hline
\end{tabular}%
}
\end{table}

Similar to case study I, we analyze the sequence of behaviors through a transition matrix. Table \ref{tab:transition_mat_case_II} shows the probability of transitioning between different driving behaviors. As depicted, when drivers are in free flow driving, there is a much higher probability to switch to a normal brake as compared to a harsh brake (0.16 vs. 0.02). This might imply that drivers are more likely to continue a more gentle driving style if they are in free flow driving which has an acceleration very close to zero. However, once being in an accelerating behavior, the driver is more likely to switch to a harsh brake as compared to a normal brake (0.15 vs. 0.13). Among the 12 participants, this might imply that an accelerating behavior continues with a harsh decelerating behavior, which can be considered a more risky driving style with high negative/positive values of acceleration. 

\begin{table}[]
\caption{Transition Matrix between different driving behaviors detected through the case study II dataset}
\label{tab:transition_mat_case_II}
\resizebox{0.49\textwidth}{!}{%
\begin{tabular}{l|llll}
\textbf{Driving Behavior} &
  \textbf{\begin{tabular}[c]{@{}l@{}}Harsh \\ Brake\end{tabular}} &
  \textbf{\begin{tabular}[c]{@{}l@{}}Free Flow\\ Driving\end{tabular}} &
  \textbf{\begin{tabular}[c]{@{}l@{}}Normal \\ Brake\end{tabular}} &
  \textbf{Accelerating} \\ \hline
\textbf{\begin{tabular}[c]{@{}l@{}}Harsh \\ Brake\end{tabular}}      & 0.75 & 0.1  & 0.12 & 0.12 \\
\textbf{\begin{tabular}[c]{@{}l@{}}Free Flow\\ Driving\end{tabular}} & 0.02 & 0.65 & 0.16 & 0.16 \\
\textbf{\begin{tabular}[c]{@{}l@{}}Normal \\ Brake\end{tabular}}     & 0.15 & 0.14 & 0.60 & 0.11 \\
\textbf{Accelerating}                                                & 0.15 & 0.13 & 0.08 & 0.64
\end{tabular}%
}
\end{table}

\begin{figure}
\begin{center}
  \includegraphics[width=\linewidth]{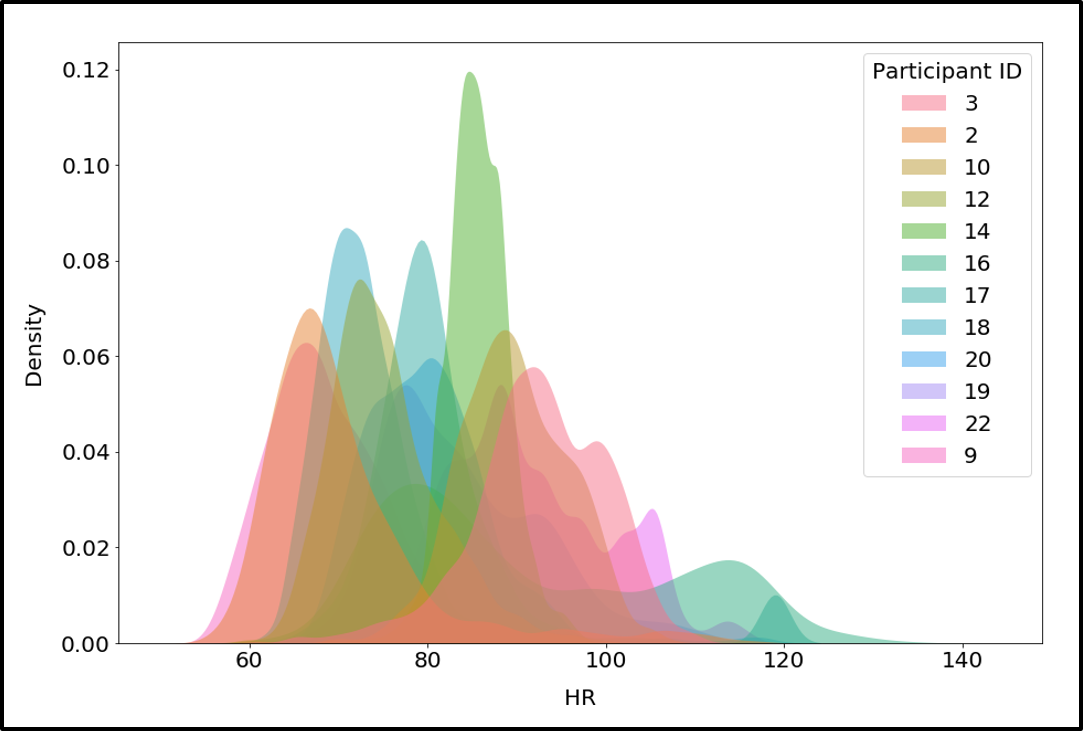}
  \end{center}
  \caption{The distribution of HR for 12 participants. Note the differences across the distributions pointing to different baselines for different participants.}
  \label{fig:hr_dist}
\end{figure}

For state data, we first normalized each participant's HR and gaze entropy values to have a 0 mean and standard deviation of 1. This is performed due to the fact that the abnormal and normal HR as well as GTE values across people can be quite different. For instance, a person's normal HR might be 65 bpm while for another person, it can be 75 bpm. For a better illustration, distribution of HR data is also shown on Fig. \ref{fig:hr_dist} where different participants' HR distribution can be very different on the mean and standard deviation, implying different normal HR values across different people.

\begin{figure}
\begin{center}
  \includegraphics[width=\linewidth]{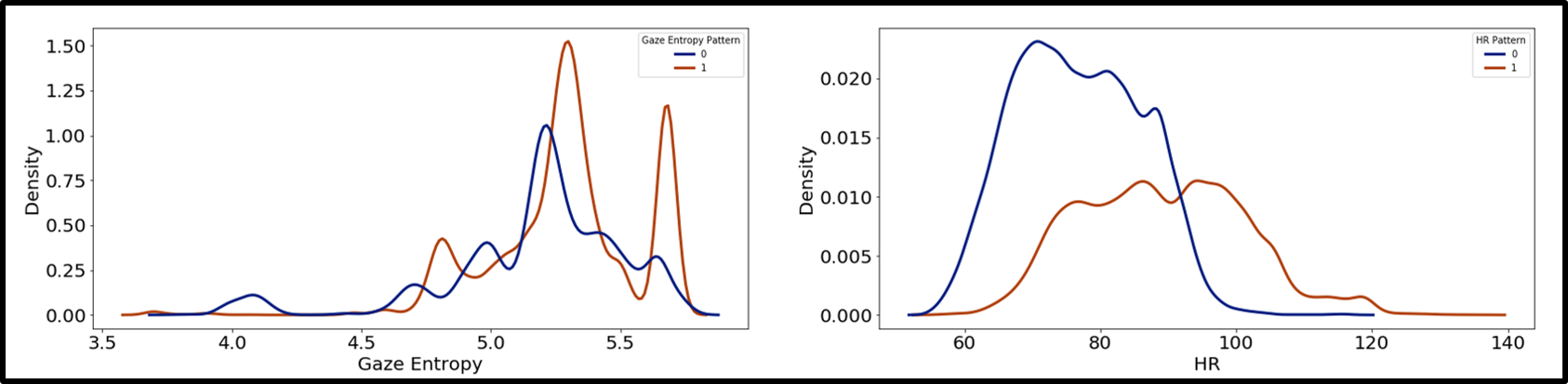}
  \end{center}
  \caption{The distribution of gaze entropy and HR for 12 participants. Note the differences across the distributions pointing to different states for the drivers.}
  \label{fig:state_cluster_case_II}
\end{figure}

We then applied the GMM-LDA method to both HR and gaze data separately, for case study II. Fig. \ref{fig:state_cluster_case_II} depicts different driver state patterns retrieved through both HR and gaze data. Similar to case study I, HR has two patterns of normal versus abnormal. Additionally, gaze entropy has two patterns of low versus high. The patterns recognized are significantly different as tested through the Kolmogorov-Smirnov test \cite{massey1951kolmogorov}. The results of the tests are depicted on Table \ref{tab:stat_test_state_case_II}.

\begin{table}[]
\caption{Detailed of statistical tests across different patterns of HR and gaze}
\label{tab:stat_test_state_case_II}
\resizebox{0.49\textwidth}{!}{%
\begin{tabular}{ccccc}
Driver State Data & Comparison & KS Test Statistic & p-value           & Significant at 0.05? \\ \hline
HR     & 0\&1       & 0.8               & \textless{}0.0001 & Y                    \\ \hline
Gaze   & 0\&1       & 0.9               & \textless{}0.0001 & Y                    \\ \hline
\end{tabular}%
}
\end{table}

\begin{figure}
\begin{center}
  \includegraphics[width=\linewidth]{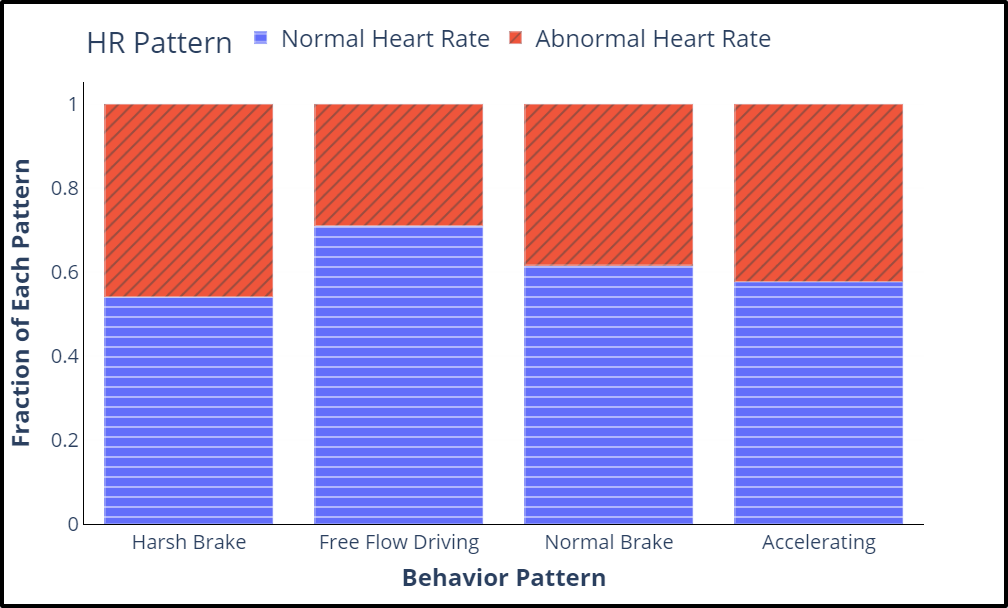}
  \end{center}
  \caption{The fraction of each HR pattern within in each driving pattern from case study II. Note that free flow driving has the highest fraction of normal HR as compared to other driving behaviors, implying a calmer state for drivers.}
  \label{fig:hr_co_II}
\end{figure}

The two detected patterns in drivers' HR can be translated into a calm and stressful state for the drivers. When considering their interaction with the driving behaviors, we observe that, harsh brakes include a higher amount of abnormal patterns in HR in our dataset as compared to normal brakes (Fig. \ref{fig:hr_co_II}). This is then supported through a Kruskal-Wallis test \cite{kruskal1952use} with a $statistic = 219.68$, and a $p-value<0.00001$. Additionally, the free flow driving with close to zero acceleration has the higher amount of normal HR, when compared to accelerating behavior. This is confirmed through a Kruskal-Wallis test \cite{kruskal1952use} with a $statistic = 71.79$, and a $p-value<0.00001$.

\begin{table}[]
\label{tab:stats_states_case_II}
\caption{Detailed statistics of HR and gaze for each recognized driver state pattern. Note that the values are normalized}
\resizebox{0.49\textwidth}{!}{%
\begin{tabular}{llcc}
\hline
Driver State Data &
  Statistical Index &
  \multicolumn{1}{l}{\begin{tabular}[c]{@{}l@{}}State Pattern 0\\ (Normal HR/\\ Low Gaze Entropy)\end{tabular}} &
  \multicolumn{1}{l}{\begin{tabular}[c]{@{}l@{}}State Pattern 1\\ (Abnormal HR/\\ High Gaze Entropy)\end{tabular}} \\ \hline
\multirow{5}{*}{HR} & Mean               & 76.78 & 89.15 \\
                            & Standard Deviation & 9.26  & 12.02 \\
                            & 25th percentile    & 69.45 & 79.76 \\
                            & Median             & 76.32 & 88.92 \\
                            & 75th percentile    & 84    & 98    \\ \hline
\multirow{5}{*}{Gaze}       & Mean               & 5.17  & 5.28  \\
                            & Standard Deviation & 0.35  & 0.28  \\
                            & 25th percentile    & 5.01  & 5.14  \\
                            & Median             & 5.21  & 5.29  \\
                            & 75th percentile    & 5.39  & 5.44  \\ \hline
\end{tabular}%
}
\end{table}

\begin{figure}
\begin{center}
  \includegraphics[width=\linewidth]{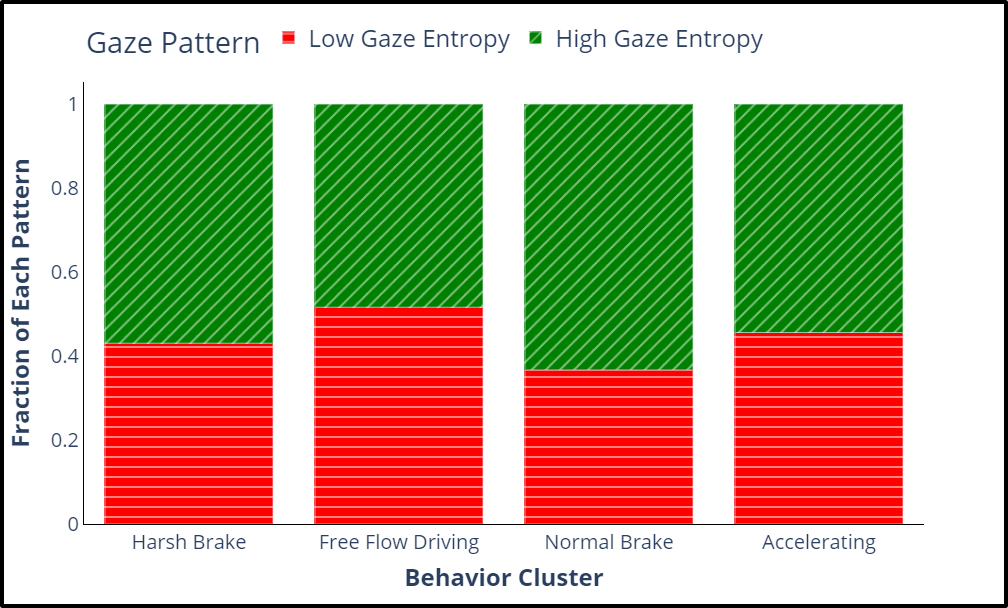}
  \end{center}
  \caption{The fraction of each gaze pattern within each driving pattern from case study II. Note that free flow driving has the highest fraction of low gaze entropy depicting lower work load. }
  \label{fig:gaze_co_II}
\end{figure}

With respect to drivers' gaze patterns, we observe that free flow driving has the lowest amount of high GTE values when compared to the accelerating behavior ($statistic = 363.390$, and a $p-value<0.00001$). This may point to the lowest cognitive load in free flow driving with close to zero acceleration as compared to the accelerating behavior with positive high acceleration (Fig. \ref{fig:gaze_co_II}). Additionally, we observe that similar to case study I, normal braking behavior has a higher GTE fractions as compared to the harsh braking behavior ($statistic = 158.969$, and a $p-value<0.0001$).

\begin{figure}
\begin{center}
  \includegraphics[width=\linewidth]{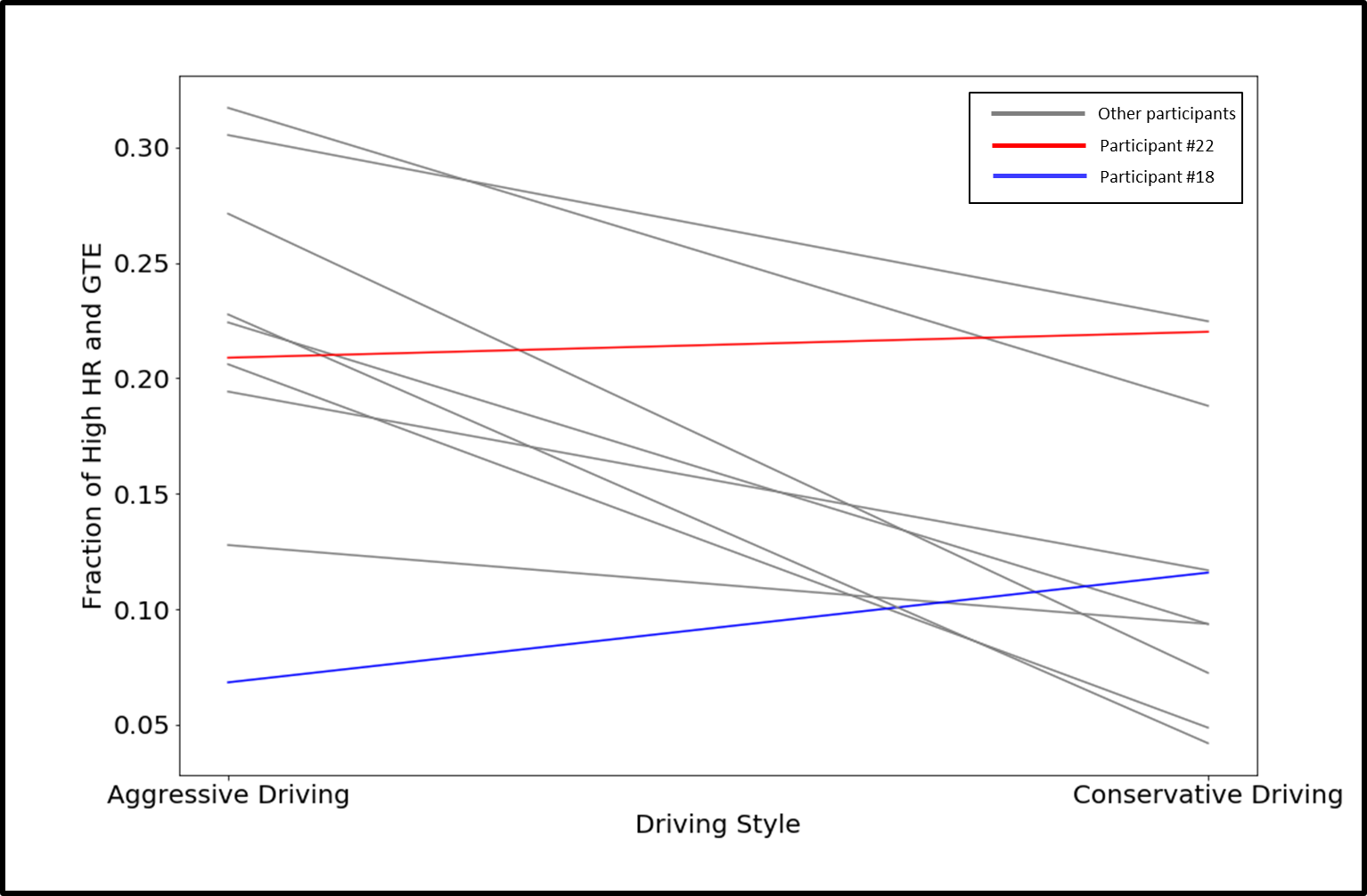}
  \end{center}
  \caption{The fraction of each driver states patterns of high GTE and abnormal HR in each of conservative and aggressive driving styles. A conservative driving style tend to keep participants HR at a calmer state while using less cognitive resources. Two of the participants do not follow the general trend shown with red and blue.}
  \label{fig:driving_style}
\end{figure}

Lastly, we aggregate the driving patterns with significantly higher than zero acceleration (both positive and negative) into one group titled as aggressive driving style. Additionally, we consider the free flow driving pattern as a conservative driving style where accelerations are close to zero. We then compare the fraction of abnormal HR and high GTE in each of these two categories of driving styles. Fig. \ref{fig:driving_style} shows the fraction of driving segments with abnormal HR as well as high GTE together with the driving styles. As shown on the graph, almost all of the drivers are less likely to have high HR as well as high GTE in conservative monotone driving as compared to aggressive high acceleration driving. We also observe that the difference between fraction of High HR and GTE between aggressive and conservative driving varies across participants leading to different slopes for the lines connecting them. Additionally, among our participants, we observe that two of them has a higher fraction of abnormal HR and GTE in conservative driving (shown with red and blue color on Fig. \ref{fig:driving_style}). The implications of these differences will be discussed later.  

\section{Discussion}
The fast-paced improvement in the development and testing of AVs has been one of the important advancements of the automobile industry over the recent years. To enhance the human-AV collaboration, realistic human behavior models are needed where a driver's comfort and behaviors can be modeled with a multidimensional approach. This is because human behavior is dynamic and can be affected by different internal dynamics (e.g., emotion and cognition), as well as external situational factors (e.g., traffic density or roadway conditions). The proposed framework in this paper couples both internal and external factors, namely context, to understand the pattern in driver's state in each driving behavior through unsupervised modeling techniques. While supervised learning has been an impactful approach over the past decade of driving research, the massive amount of data collected through past NDS points towards more efficient, fast-paced methods that can understand behaviors with no or minimum manual annotation. While in the presented case study I, we have used very high-resolution data from one participant for method illustration, case study II confirms that low-resolution data when coupled with unsupervised modeling techniques, can help detect certain states and behaviors across participants. As current vehicles are equipped with multiple sensors such as GPS, our case study II suggests that coupling a conventional wearable device with sensors that are already in vehicles can provide significant insight into the driver's state in each driving behavior.   

Driver's state patterns indicated the level to which a driver might be affected by different driving behaviors. More specifically, in our first case study, using high precision data, we observed that the driver had a higher fraction of normal HR in free-flow driving on the highway, implying a calmer state for the driver. This has then been confirmed in our case study II where we analyzed the data from 12 participants. Free flow driving not only had the lowest fraction of high gaze entropy, implying a lower workload but also was accompanied by the least fraction of abnormal HR among our participants in case study II. These findings are in line with previous studies indicating less reported subjective emotional states \cite{dittrich2021drivers}, as well as calmer states \cite{tavakolipersonalized} in highway driving. Additionally, our results suggest that even the same action can have different effects on the driver when performed with different styles. For instance, we observe that harsh brakes are more likely to be accompanied by abnormal HR values implying higher stress levels as compared to normal brakes.

Our method can be leveraged to understand the effects of different driving styles on drivers' states. Multiple studies have pointed out the importance of driving style selection for drivers' trust in AVs \cite{ekman2019exploring}. For instance, a recent study points out that a ''defensive'' driving style as compared to an aggressive driving style is perceived to be more trustworthy \cite{ekman2019exploring}. In case study II, our results suggest that a more conservative driving style that is accompanied by close to zero acceleration values is less probable to be accompanied by lower workload and stress levels. Although our sample size might not generalize to the larger pool of drivers, it shows the importance of choosing driving styles according to the drivers' psychophysiological states given specific road environments (e.g., highway versus city driving). For instance, our results show that free flow driving is a better fit for the drivers of our study in keeping their HR at a normal pattern. Choosing highways over crowded city streets can help driving with the free flow speed, which then helps to keep drivers calmer with a lower workload. These differences in stress levels and workload have strong implications for humanizing automated services, including both the AV and routing services. Taking our results into account, different driving behaviors can be chosen based on how to they might affect a driver based on the unsupervised modeling of their historical driving data within the safety boundary of each behavior. Additionally, services such as route selection can take these factors into consideration for providing a human-centered service that considers how a driver might feel through driving on each route based on potential driving behaviors that can occur on the route (e.g., amount of highway driving, curve driving, etc.). 

When comparing conservative versus aggressive driving style, we observe different levels to which participants are affected by each style, which might be indicative of individual differences when being in the two different driving styles. Additionally, among our participants, we have seen two participants that had higher abnormal HR and GTE during conservative driving style. This was in contrast with all the other 10 participants. After careful examination of the videos and interviewing the participants, we realized that one of the two drivers uses loud music in highways which might be indicative of the high HR values in conservative highway driving scenarios (shown with red on Fig. \ref{fig:driving_style}), while, the second driver often uses cruise control in highway driving (shown with blue). Although more data is required to address the effect of loud music in naturalistic driving, part of our future work will be focused on using the data provided in HARMONY \cite{tavakoli2021harmony}, to assess the effect of music in naturalistic scenarios. This preliminary finding paves the way for understanding the naturalistic effect of music on driving behaviors and driver's states.

Additionally, while we did not have access to the cruise control usage data by the vehicle, we hypothesize that the abnormal HR and GTE state for the participant shown with blue on Fig. \ref{fig:driving_style} could be due to interaction with the semi-automated cruise control system. Current cruise control systems are generally used to keep a constant speed during highway free flow driving. This might imply that although automated driving assisting systems (ADAS) can help the drivers for delivering a safe driving task, they might be the cause of more stress and might result in changes in drivers' state that can only be detected through using psychophysiological metrics. Changes in drivers' behaviors and states, while engaged in ADAS systems were mentioned previously in other naturalistic driving study researches (\cite{dunn2019understanding,kim2021driving}). It is possible that when the cruise control system was active, which is often on highways and at higher speeds, the driver had more stress as they had to supervise the automation. Alternatively it could also be the case that participant engaged in other activities or started mind-wandering, which increased the fraction of high HR and GTE. A very recent study focusing on semi-automated vehicles drivers is also pointing towards more eyes-off-the-road during using automated systems \cite{morando2021model}. More research with taking psychophysiology into account is required to understand the real effect of shared-autonomy on the driver. This also has strong implications for human sensing modules inside vehicles that rely on computer vision. Such modules often only consider the driver's facial expressions and gaze patterns while leaving behind the driver's physiological responses such as HR, skin temperature, etc. While computer vision has shown to be a very strong method for detecting a driver's state, coupling it with physiological metrics can provide a deeper understanding of the driver's state.  

Although we have not focused specifically on the sequence of behaviors, using Tables \ref{tab:transition_matrix} and \ref{tab:transition_mat_case_II} we observe important insights regarding the transition probabilities across different behaviors. This helps with analyzing driving styles in a deeper fashion. From Table \ref{tab:transition_matrix}, we first observe that a harsh braking behavior is often followed by a normal braking behavior, while a normal braking behavior is more likely to continue in normal braking. This might imply that while performing normal braking, the driver is more confident regarding the context such as the surrounding road environment, traffic density, or other vehicles, thus the chances of switching to a harsh brake is lower than other two behavior. We also observe that the braking behavior that succeeds free flow driving is divided between the two types of braking behaviors. This might imply that depending on the context driver chooses between two different behavioral patterns. More information from the context can provide insight into the reasoning behind the choice between the two. Information such as the presence of other vehicles, the speed of other vehicles, and their distance can be predictive of such choice. This will be analyzed in our future work. Additionally, from Table \ref{tab:transition_mat_case_II} we observe that when being in accelerating driving behavior, the probability of switching to a hard brake is the highest among other behaviors. This indicates that aggressive, less monotone driving through accelerating behavior is more likely to result in a harsh braking/deceleration behavior, which can be inferred as continuing the less monotone aggressive driving. On the other hand, being in a free flow driving behavior is much less likely to switch to a harsh brake ($p = 0.02$). Such results could indicate the persistence of both aggressive or conservative behavior throughout the time, which can be used for driving style prediction. One part of the future work will be focused on this area to understand and predict the sequence of behaviors in different contexts.  

Lastly, the findings of our approach to driver state modeling when applied in real-time, has the potential to guide an autonomous vehicle to better predict human behavior and take actions that are fit to driver's physiology. Note that a vehicle does not need to know the meaning of each unsupervised pattern. For our study, we translated driving patterns into driving behaviors, which helps with illustrating each pattern and justifying the method. However, a vehicle only requires to know the characteristics of each planned driving pattern, compare it with the detected patterns in the historical data, and the probability of abnormal HR within different planned behaviors. This can then help the vehicle choose among different actions. For instance, when choosing between a harsh and normal brake, the vehicle can estimate the acceleration in each action, find the most similar driving pattern to each action based on acceleration characteristics and model fit, and choose the one that is more suitable to drivers' HR and gaze patterns. Additionally, using our approach in driver state modeling, we can better predict future driving decisions by building behavioral models of past driver states and incorporating theoretical findings in psychology such as memory based judgment and decision making \cite{sharif2016effect,sharif2021effect}.

\section{Conclusion and Future Work}
In this paper, we propose a method to understand driver's reactions to different driving maneuvers in real-world driving scenarios by using unsupervised methods. Using our approach, different driving behaviors can be ranked based on how they affect a user's well-being, thus producing actions that are more aligned with human's state, such as stress levels and cognitive loads. Our method uses a combination of Bayesian Change Point detection together with GMM-LDA methods applied on both driving and human sensing data to mine such associations between driver's state and behavior. 

Our work has five main limitations, which will be addressed in the future work of this paper. First, we will increase the number of participants as well as the duration of collected data to understand individual differences in how various driving styles may affect each driver. This can help us cluster different drivers based on how they are affected by the driving behaviors. Second, we will perform deeper feature extraction on both driver's HR as well as gaze variation to retrieve a deeper understanding of the driver's state in each driving pattern. Third, we will perform the same analysis with a higher number of sensors and modalities in both the human sensing and vehicle sensing part of our study. This can include modalities such as driver's other bio-signals (e.g., skin temperature) and modalities retrieved from the computer vision techniques (e.g., distance to other vehicles). Fourth, we will focus deeper on the sequence of driving behaviors to be able to predict behaviors using time series prediction methods. Lastly, we will consider other unsupervised methods that can help understand behaviors in a more fine-grained manner.

\section{Declaration of Competing Interest}
The authors declare that they have no known competing financial interests or personal relationships that could have appeared to influence the work reported in this paper.

\section{Acknowledgement}
We would like to thank the UVA Link Lab, and Commonwealth Cyber Initiative (CCI) for providing support and resources to enable this project. Also, we are thankful to the UVA Institutional Review Board for their continuous support and feedback.

\bibliographystyle{IEEEtran}
\bibliography{bib}

\begin{IEEEbiography}[{\includegraphics[width=1in,height=1.25in,clip,keepaspectratio]{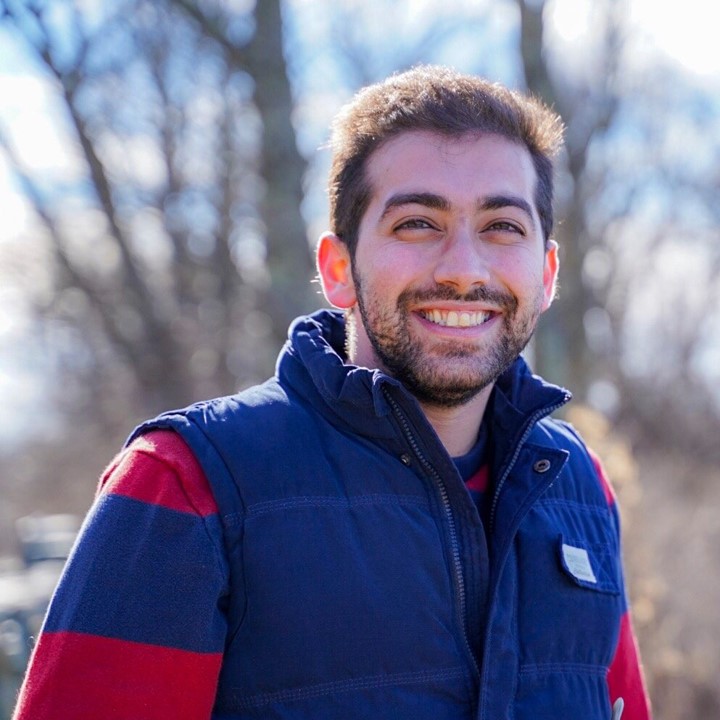}}]{Arash Tavakoli} Arash Tavakoli is a PhD student in the Engineering Systems and Environment department as well as the Link Lab at the University of Virginia. He has earned his BSc. and MSc. in Civil Engineering from the Sharif University of Technology and Virginia Tech, respectively. Arash’s research interest lies on the intersection of transportation engineering, computer science, and psychology.
\end{IEEEbiography}

\begin{IEEEbiography}[{\includegraphics[width=1in,height=1.25in,clip,keepaspectratio]{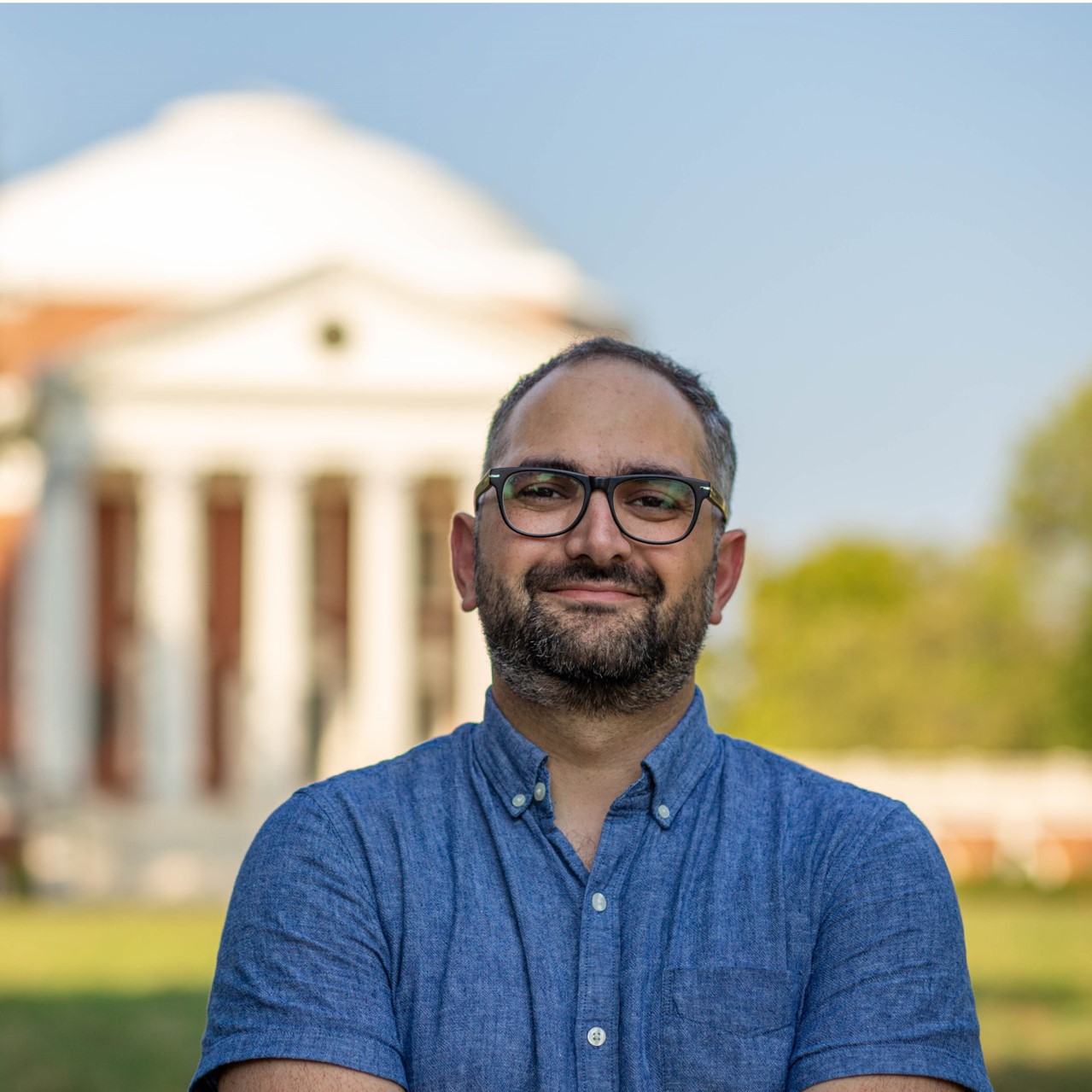}}]{Arsalan Heydarian} Arsalan Heydarian is an Assistant Professor in the department of Engineering Systems and Environment as well as the UVA LINK LAB. His research focuses on user-centered design, construction, and operation of intelligent infrastructure with the objective of enhancing sustainability, adaptability, and resilience future infrastructure systems. Specifically, his research can be divided into four main research streams: (1) intelligent built environments; (2) mobility and infrastructure design; (3) smart transportation and (4) data-driven mixed reality. Dr. Heydarian received his Ph.D. in Civil Engineering from the University of Southern California (USC), M.Sc in System Engineering from USC, and B.Sc. and M.Sc in Civil Engineering from Virginia Tech. 
\end{IEEEbiography}

\end{document}